\documentclass[aip,jmp,amsmath,amssymb,
,]{revtex4-1}
\pdfoutput=1

\usepackage{graphicx}% Include figure files
\usepackage{dcolumn}% Align table columns on decimal point
\usepackage{bm}% bold math

\usepackage[mathlines]{lineno}% Enable numbering of text and display math

      \usepackage{color}
      \usepackage{amsfonts}
      \usepackage{mathptm}

       \def\z0{z_{0}^{\tiny (l)\normalsize}}
       \def \be{\begin{equation}}
       \def \ee{\end{equation}}

       \newcommand{\mm}{\mathrm}

       \newenvironment{narrow}[2]{%
         \begin{list}{}{%
             \setlength{\topsep}{0pt}%
             \setlength{\leftmargin}{#1}%
             \setlength{\rightmargin}{#2}%
             \setlength{\listparindent}{\parindent}%
             \setlength{\itemindent}{\parindent}%
             \setlength{\parsep}{\parskip}}%
         \item[]}{\end{list}}
\begin{document}

%\preprint{}


\title{Non-perturbative features of driven scattering systems}%

\author{Andrea Cintio}
\affiliation{Department of Physics, University of Pisa, 56127 Pisa Italy}
%\affiliation{Max-Planck-Institut f\"ur Plasmaphysik, Euratom Association, D-85748 Garching Germany\\ in collaboration with Max-Planck-Inst. f\"ur Sonnensystemforschung, Katlenburg-Lindau Germany}

%\author{Liu Chen}
%\affiliation{Institute for Fusion Theory and Simulation, Zhejiang Univ., Hangzhou, People's Rep. of China\\
%and Dept. of Physics and Astronomy, University of California, Irvine, CA 92697-4575 USA}%

\author{Fulvio Cornolti}%
\affiliation{Department of Physics, University of Pisa, 56127 Pisa Italy}%

%\author{Fulvio Zonca}
%\affiliation{Associazione Euratom-ENEA sulla Fusione, C.R. Frascati, C.P. 65 - 00044 Frascati Italy.}%


%%\author{}
%\thanks{}
%%\affiliation{}


\date{\today}

\begin{abstract}

     We investigate the scattering properties of one-dimensional, periodically and non-periodically forced oscillators. 
% We are interested in their dynamics, close to the heteroclinic orbit, in a part of the phase space where KAM scenario is absent.
     The pattern of singularities of the scattering function, in the periodic case, shows a characteristic hierarchical structure where the number $N_c$\ of zeros of the solutions plays the role of an 
     order parameter marking the level of the observed self-similar structure. The behavior is understood both in terms of the return map and of the intersections pattern of the invariant manifolds of the outermost
     fixed points. 
In the non-periodic case the scattering function does not provide a complete development of the hierarchical structure. 
%some orbits, that experiences very long motions close to the parabolic orbits, are lost.   
The singularities pattern of the outgoing energy as a function of the driver amplitude is connected to the arrangement of gaps in the fundamental regions. The survival probability distribution of 
     temporarily bound orbits  is shown to decay asymptotically as a power law. The ''stickiness'' of regular regions of phase space, given by KAM surfaces and remnant of KAM curves, is responsible for 
     this observation.   
%     We find that, in general, there are regions of the phase space where a variable which refers to the asymptotic motion sensibly depends as on initial conditions, given the external force, and as on the latter given
%     the former. Differences between periodic and nonperiodic cases are found about the location of the phase space regions with irregular behavior. A quantitative analysis of irregular dynamics is also performed by means 
%     of the Lyapunov's coefficients.

\end{abstract}

\pacs{05.45.+b, 03.20.+i}
%\keywords{}

\maketitle

\section{Introduction}

   In a scattering problem the system is characterized by having a localized region of configuration space ('interaction region') where the interaction is relevant, while elsewhere the interaction is absent. A particle is injected into the interaction region and experiences a scattering event; after each event it may  escape from the interaction region forever ('scattered' trajectories) or eventually proceed to another scattering event. A map that links some initial parameter to some feature of the outgoing state is called a scattering function. There may exist 'trapped' trajectories, i.e. particles that wander for ever in the interaction region. In these circumstances the scattering function is not defined.  
% An irregular dynamics, that is "infinitely sensitive" (singular) on a (fractal) set of either initial conditions or system parameters, is not expected in a scattering process, i.e. in the case of motion in an unbounded phase space (\cite[]{DF-1976}).\\ 
%   It is possible to discretise the dynamics of a single scattering event by a map that describes in the phase space how incoming parameters (incoming state) for the present single scattering change into the parameters 
%   that are either the incoming parameters of the next scattering, if it occurs, or the parameters of a scattered orbits when the particle reaches an asymptotically free region.  
A scattering process is called 'irregular' or 'chaotic' when a scattering function is singular on a fractal set of variable initial values (see Ref.~\onlinecite{Eckhardt-1988}).
%   Let us consider a map that links the state before the scattering (initial state) to the asymptotic one after the scattering  (final state); in some regions of the phase space it can provide an irregular behavior.\\ 
%   The transition from a regular interval to an irregular one is not smooth, namely the map can only have either regular or irregular features.

A condition for having chaotic scattering is the existence of an invariant set\ $\Lambda$\ of an infinite number of unstable, periodic and aperiodic orbits, localized in the interaction region. Whenever $\Lambda$\ contains a Cantor set, irregular scattering occurs. 
%we can observe an irregular dynamics also in  case of a motion in an unbounded phase space (\cite{JS-1987}). \\
   In fact the stable manifolds of orbits in $\Lambda$ reach out into the asymptotic region by means of the Hamiltonian flow; an orbit started on one of the stable manifolds is captured in the interaction region for all the time, i.e. the final state is not defined. Hence there exists a fractal set of asymptotic in-variables (the initial conditions lie in the stable manifolds of orbits in\ $\Lambda$) where the scattering function is singular \cite{JS-1987,JS-1988}. 
   The pattern of homoclinic/heteroclinc connections of invariant manifolds of $\Lambda$ is called 'chaotic saddle'.

%   whose final state is not defined because they end up in the interaction region;  we ought to consider the orbits whose initial 
%   conditions lie in the stable manifold of orbits in\ $\Lambda$.\\
   Therefore incoming parameters close to each other and close to the 'stable manifold' of the chaotic saddle can be mapped to very different outgoing parameters. They spend a long time in the vicinity of $\Lambda$\  and trace out the type of motion performed by the localized trajectories. Similarly outgoing parameters close to each other and close to the unstable manifold can come from very different incoming parameters. $\Lambda$\ acts as a ''repeller'' with global influence on the asymptotic region.
   Therefore the scattering function displays wild fluctuations on all scales on each set containing initial conditions leading to trapped orbits. The origin of the fractal singularities of a scattering function is the chaotic motion in the invariant bound set; the irregularity of the motion in an unbound phase space is connected to a form of bound chaos.
%  A stable orbit is defined here as an orbit where the final state is contained in a region near the interaction region.

%   In classical systems the irregular or chaotic scattering is a representation of the transient chaotic behaviour (\cite[]{EJ-1986}) (\textsl{prendere spunti dall'introduzione dell'articolo}).

%   The set of bound orbits forms a invariant thread-ball structure in the phase space. For every stable manifold of this type we can find an unstable manifold, where the orbits tend either to get further or closer, with 
%   an exponential law in terms of the phase space parameters. Consistently, the map linking the final states to the initial states corresponds to a singular scattering function. All orbits belonging to the stable manifolds 
%   end up in the interaction region. 

% (?) A chaotic repeller is generated by the fact that in the localized region of interaction unstable periodic orbits exist.
Thus, the main issue in studying a problem of chaotic scattering is to determine the properties of the invariant set, in particular the topological pattern of the chaotic saddle  and the character of the dynamics in it. In this regard it is hoped that an equivalent symbolic dynamics is found out to explain the behavior of a scattering function. The invariant set provides a fractal structure whose dimension, Lyapunov exponents and escape rate can be connected to the particular arrangement of the singularities in the scattering function or in the delay time function.
 
%    If we know the Hamiltonian, the main problem is to determine the chaotic invariant set. We are mainly interested in a structural understanding of the dynamics. For a system of chaotic scattering this means that is not necessary to know all the analytical details of the solutions in the chaotic set; often such an attempt is hopeless. It is more interesting to reconstruct the topological pattern of the chaotic saddle from the equations of motion (this includes the attempt to catch a symbolic dynamics) and, through it, explain the behavior of a scattering function. The invariant set provides a fractal structure whose dimension, Lyapunov exponents, escape rate can be connected to the particular arrangement of the singularities in the 'scattering function' or the 'time-delay function'. 

%    In chaotic systems we always have homoclinic and/or heteroclinic intersections of invariant manifolds. As Smale shows (\cite{smale-1967}) this is connected with the horseshoe construction. A system with 
%    $n$\ fixed points (in particular, periodic solutions) corresponds to a horseshoe with same number of fixed points
    
    The chaotic (invariant) set shows a relatively simple structure when it is completely hyperbolic. In this case it consists of unstable periodic orbits (all of them are hyperbolic) and their homoclinic and heteroclinic connections only. 
%In the non-hyperbolic case the invariant set contain KAM tori and sets of marginal stability; when we change a parameter in some range of values we may modify the estension, the 
%    shape of these structures and we may even destroy them. As a consequence the topology of the intersections and so the dynamics in the invariant set are modified.    
%    In chaotic systems we always have homoclinic and/or heteroclinic intersections of invariant manifolds. 
    As Smale shows \cite{smale-1967}, this is connected with the horseshoe construction. A system with $n$\ fixed points (periodic solutions) corresponds to a horseshoe with the same number of fixed points; one can obtain the horseshoe structure by plotting the invariant manifolds of any unstable periodic
    point. The invariant set may contain KAM tori and sets of marginal stability; in this case it is not completely hyperbolic and the horseshoe is not fully developed. We will present a system which differs from a standard Smale's horseshoe model: 
    \begin{itemize}
     \item [-] the configuration space is infinite, the interaction is long ranged and parabolic orbits exist; it follows that the invariant set is not compact and the system  is described by a symbolic dynamics with an infinite number of symbols; 
     \item [-] in the phase space there are islands which consist of unbroken KAM surfaces. 
    \end{itemize}

    Many physical scattering systems can be modeled in the following way. We consider a set $S$ in the phase space. A driving force removes a fraction of points from $S$  after each step of an iterative process: after the first step, it leaves a number of intervals in $S$, after the second step a set of new subintervals replaces each old interval. This process can be infinitely repeated. The structure of the set of points  that are never removed is Cantor-like. This process is characteristic of systems with fully developed chaos. 

In chaotic scattering the Smale's mechanism leads to an exponential decay of phase space  ensembles: all phase space points are depleted except for the invariant set, which consists of all trapped orbits of the system.
However the assignment of an exponential decay law to a chaotic dynamics is not general. When in the phase space some regular islands with KAM surfaces survive,  the particle initialized in a chaotic region can spend a long time near the boundary of a regular region. In this case, the leaving away of the particles from the KAM region gives an algebraic decay \cite{Chirikov-Shepelyanski,Karney,Meiss-Ott}.

    The existence of KAM surfaces is not the only mechanism resulting in an algebraic decay. In the Ref.~\onlinecite{Hillermeier-Blumel-Smilansky} the authors present a chaotic system that does not show any KAM structures and that decays according to a power law. The scaling is explained through a model of random walk in phase space; the dynamics of symbolic strings is Markovian. Beeker et al. \cite{Beeker-Eckelt}  observe an algebraic scaling of the scattering function.  Nevertheless they claim that the KAM component effects are negligible for the parameter values they choose. Beeker et al. provide reasons for the scaling by means of a different mechanism than the stickiness of KAM tori. They consider the survival time of the temporarily bound orbits close to the parabolic orbit. If the system is described by a symbolic dynamics with infinite symbols, one can have a chaotic dynamics in a phase space without any regular region and, at the same time, a scaling with a power law.    
           
    When an external control parameter is varied, the invariant set changes its structure. In particular, if some non-hyperbolic components enlarge, then the stable and unstable bundles can be 
    displaced one with respect to each other. For certain parameter values, intersections are destroyed and some localized orbits of the invariant set get lost, i.e. bifurcations take place. As a consequence, the invariant set may be no longer topologically equivalent to a complete Cantor set. 
%\comment{tuttavia la parte la topologia che e' rilevante per ordini più bassi della struttura non e' modificata da piccole variazioni dei 
%    parametri (non uniformita')}
    Some relevant quantities (such as the escape rate and the topological entropy) have general features: as functions of the parameter they display plateaux, 
    where no orbits disappear. There is also a set of parameter values where the quantities change indicating the presence of bifurcations. This set can have an involved structure; in fact for a class of systems the topological entropy exhibits a devil's staircase \cite{Lai-Zyczkowski-Grebogi}.

%   In classical systems the irregular (or chaotic) scattering is the Hamiltonian version of the ''transient'', chaotic behaviour (\cite[]{EJ-1986}).

%   The chaotic (invariant) set has a relatively simple structure only when it is completely hyperbolic, as in the case of the invariant set occurring in the horseshoe construction (\cite{smale-1967}). It consists of an 
%   infinite set of unstable, periodic orbits and of their homoclinic and heteroclinic connection and does not contain any KAM tori and cantori.\\ 
%   But in the non-hyperbolic case some subsets of marginal stability have to persist. 

%   The invariant set provides a fractal structure whose i.e. dimension, Lyapunov exponents, escape rate can be extracted  from scattering data, in particular from the arrangement of the singularities in the 'deflection
%   (or scattering) function' or the 'time-delay function'. 

   There is a wide variety of applications of the chaotic scattering in classical physics.
 
   In celestial mechanics complicated motions can take place for close encounters, but, after, the particles separate, with the exception of a set of 
   initial conditions of zero measure \cite{Petit-Henon}. 

   In many chemical reactions (in particular, in the case of a biatomic molecule interacting with an atom) small changes in the initial conditions lead to drastic differences in the final states so that a nonreactive 
   trajectory may exist in the vicinity of a reactive one. Pollack et al. \cite{Pechukas} emphasized the importance of unstable periodic orbits that, typically, provide a non-attracting, infinite set 
   \cite{Koch-Bruhn}. Skodje et al. \cite{Skodje} got a chaotic scattering approach to the matter.

   In hydrodynamics the scattering among more than three ideal, linear vortices can bring to chaotic dynamics. Aref and Eckhardt et al. \cite{Eckhardt-Aref, Aref} draw attention to the processes occurring in the 
   interaction. Two couples of vortices  that interact can either pass by each other or exchange their partners and then spend time moving along circular motion, until the next collision when they can either 
   convert to the original couples or scatter away from each other. This capture state is unstable and exhibits all properties of chaotic scattering, and, in particular, it may be disrupted by a 
   small change in the parameters.

   There are many applications of the chaotic scattering to the classical scattering problem  by a potential. The heart of all examples is the existence of unstable, bounded orbits that trap a set of orbits in the 
   continuum of the scattering trajectories and lead to singularities in the scattering functions \cite{EJ-1986,Jung-1986,Jung-1987,JS-1987}. A particular class of problems is connected with the internal dynamics  of atomic systems or clusters interacting with an external electromagnetic field. Even if the motion is integrable in the field-free case, external driving 
   generically destroys integrability and can lead to irregular dynamics, sensitive with respect to both initial conditions and parameters of the driver. This class of systems contains models of one-dimensional oscillating
   potential wells.
 
   In this paper we deal with a representative of this class of systems: the model of 'rigid spheres'  \cite{Mulser,Kundu-Bauer}. It is introduced in connection with the problem of the energy absorption in clusters irradiated by intense 
   electromagnetic fields; the model provides a basic mechanism of absorption. We focus on some features of the dynamics of the model. These features are understandable by means of a chaotic scattering approach.

In section 2, we introduce the model and we list the particular cases we consider in the numerical study. In section 3, we summarize the relevant numerical results about the scattering function and its structure of singularities; the geometry of the scattering function is dominated by the escape orbits while the
    singularities are caused by the captured orbits. We show that the scaling of the delay time function is algebraic. 
    In section 4, following  Alekseev  \cite{alekseev,alekseev-3} and Ref.~\onlinecite{Eckelt-Zienicke} we introduce a return map and we obtain a detailed description of the structure of the scattering function. 
    In section 5, we give a different approach to the investigation of the structure of the scattering function. We construct the pattern of intersections of invariant manifolds of the two outermost fixed points. The 
    singularities structure of the delay time function corresponds to the pattern in which the bundles of stable manifolds intersect the local segment of the unstable manifold. 
    As Smale has shown, the homoclinic/heteroclinic bundle is connected with a horseshoe construction. We obtain that the horseshoe map is not completely developed due to the non-hyperbolic effects created by the surfaces of KAM islands and their secondary structures; they cause the algebraic behavior in the statistics of the delay time function. 
    In section 6, we investigate the outgoing energy as a function of the driver amplitude for fixed initial conditions. We observe that this function displays behaviors similar to the scattering function. This is interpreted through some essential properties of the topology of the homoclinic/heteroclinic bundle.  The discrepancies between the two functions depend on secondary aspects of the topology.  

    \section{The problem}
 We study the scattering of a particle (mass m=1) by a one-dimensional potential well $V(x)$,\ $x\in\mathbb{R}$,\ driven by a homogeneous, time dependent force\ $f$.

The potential we consider is attractive, bell-shaped, symmetric, limited, with a single equilibrium point. The definite expression we treat is provided by the model of rigid spheres. It is a simple electrostatic system where a negatively charged sphere moves in the potential generated by a fixed, positively charged one. The potential is Coulomb-like when the spheres do not overpose; otherwise it is polynomial.  Following Ref.~\onlinecite{Kundu-Bauer} we assume      
        \be \label{eq:oscillatore-polinomial}
          V_1(x)\ =\left \{ \begin{array}{lc}
                                 1/2\,x^2\,-\,3/16\,|x|^{3}\,+\,1/160\,|x|^5\,  & |x|\le 2 \\
                                 -1/|x|\,+\,6/5  &  otherwise
                              \end{array} \right .
      \ee
      $V_1$\ is $\mathbb{C}^2$\ for\ $x\in \mathbb{R}$.

For reference, we also perform simulations with a Lorentz-shaped  potential  $V_2$\ characterized by general features similar to  $V_1$ except for being  $\mathbb{C}^{\infty}$. 
         \be\label{eq:oscillatore-lorentz}
           V_2(x)\ =\ \frac{6/5\ x^2}{(12/5\,+\,x^2)}
         \ee
We also test potentials of different shapes with the same general features: the results relevant for our discussion are essentially the same.

The system is driven by a homogeneous, time-dependent force $f$ representing the effect of an applied oscillating electric field. We study two cases: $f=f_1$ 
       (compact support) and $f=f_2$ (periodic).
      \be\label{def:gtn}
       f_1(t)\ =
       \,\left\{\begin{array}{lc}   
       e_0\sin^{2}(\frac{\nu\,t}{n})\ \,\cos(\nu\,t) & 0\le \nu\,t\le n\,\pi \\
%      \sin^{2}(\frac{\nu\,t}{m}\,+\,\frac{(m-n)}{m}\,\frac{\pi}{2})\ \,\cos(\nu\,t) & \frac{n\,\pi}{2}\le \nu\,t\le \frac{m+n}{2}\ \pi \\
       0 & otherwise
      \end{array}
      \right .
      \ee
$n$\ in $\mathbb{N^+}$\, $n>>1$. The support of the envelope is one period of $\sin^2(\frac{\nu\,t}{n})$; in this way it is introduced a time scale $n/\nu$\ . Therefore $f_1$ is defined by the amplitude $e_0$, the width of the envelope and the frequency $\nu$\ of the co-sinusoidal factor.\\      
The second form of $f$ we study is purely sinusoidal
      \be\label{def:gtnX}
       f_2(t) =e_0\sin(\nu\,t)
     \ee
We always assume $\nu > 1$.

      The time-dependent Hamiltonian is given by
      \be\label{def:hamiltonian}
       \begin{split}
        H(p,\,x,\,t)& =\,H_0(p,\,x)\ -\ e_0\ f(t)\ x\\ 
                    &=\ p^2/2\ +\ U(x,\,t)           
       \end{split}
      \ee     
      where\ $H_0(p,\ x)=\,p^2/2\ +\,\omega^2\ V(x)$, and, according to the case,\ $V=V_i$ and $f=f_i$\,,\ $i=1,2$.      
      In both cases the undriven oscillators are characterized by a linear angular velocity $\omega=0.7$ and  an escape solution $x_h$ ('heteroclinic' orbit) with parabolic critical points, i.e.  $|x_h(\pm \infty)|=\infty$\ and\ $\dot{x}_h(\pm\infty)=0$.  The energy of the unperturbed escape orbit is $6/5\,\omega^2=0.588$\,.

The phenomenology we study concerns drivers with strength even many times the maximal oscillator force. Therefore it cannot be fully understood only through perturbative 
 methods. However there are some perturbative treatments (in particular  \cite{Moser-73,Cicogna-01,Dankowicz}) that show the important role of the escape orbit in the 
explanation of the irregular dynamics of the driven oscillators. Thus 
%We expect that our approach should focus on a non-perturbative mechanism as we deal with properties that take place in a large range of the driver amplitude.\\
% The points that should be relevant are a) the oscillator possesses only a finite interval where the inner force is not negligible with respect to the driver (interaction 
%region) b) the oscillator possesses some heteroclinic orbits whose critical points lie far away the interaction region.
we concentrate on orbits in the phase space with energies not far from the escape energy of the free oscillator. In such situations, at fixed strength of the driver, in the phase space there are size-able regions of initial conditions characterized by unstable trajectories which appear to be so erratic that the system motions can be regarded as stochastic. Elsewhere, irregular motions are absent. In addition, if the strength of the driver is varied then stochastic regions move in the phase space. 
%The phenomenology concerns drivers with strength also many times the maximal oscillator force. Therefore it cannot be fully understood only through perturbative methods.\\

The properties of the dynamics can be approached by studying the scattering functions, i.e. maps connecting the initial states to the final ones, after a time long enough such that the particle can move in the interaction region and possibly comes out of it. 
In order to determine the scattering functions, we are interested in asymptotic data. For the case $f=f_1$ we get the relevant quantities at $t\ge n\,\pi/\nu$. The driver should play the role of a transient. Actually, we find that the main descriptive properties are developed, at least partially, before the end of the transient. In the case  $f=f_2$ we simply take the data after many periods of $f_2$. 

The maps we obtain display the irregular behavior of the dynamics: intervals of the initial state variables, where the map is sensitively depending on, alternate with intervals where the map is smooth.
 
The irregular behavior of the scattering functions can be analyzed through the approach of the irregular scattering theory.   
Actually the chaotic scattering theory deals with physical situations where the dynamics takes place in an infinite volume of the phase space, the interaction is confined to a finite region and there are some scattering  functions which display wild fluctuations on all scales.

 %    Numerical calculation are performed on two systems. We integrate the equation of motions and obtain some discrete dynamical maps.

%      We are interested in checking whether analogous features are observed in other oscillators.  
%      Let us note 
%      \[V(x)=1/(2\,\pi)\int_0^{2\,\pi/\nu}dt\ U(x,\,t)\] 
%      The parameter $\omega$\ is the proper frequency of the oscillators $V$, after that it is linearized close to the equilibrium point. In this paper we set $\omega =0.7$\\ 
%      The parameter $e_0$\ controls the strength of the time dependent part of $U(x,\,t)$; $e_0=0$\ corresponds to the undriven oscillator.
      
%      In both cases the undriven oscillator possesses an escape solution $x_h$ ('heteroclinic' orbit) with parabolic critical points, i.e.  $|x_h(\pm \infty)|=\infty$\ and\ $\dot{x}_h(\pm\infty)=0$.\\ 
%      The energy of the unperturbed escape orbit is $6/5\,\omega^2=0.588$.

      We briefly give an outline of the main results of the numerical calculations. 
      \begin{itemize}
       \item [-] For both the oscillators with the sinusoidal driver $f_2$ we determine two scattering functions.
            For a trajectory with initial velocity $\mathrm{v}_0$\ we get the energy $H_0$ and the number $N_c$\ of crossings through $x=0$\ after the particle has moved out from the interaction region.\\ 
            Then we plot $H_0$\ and $N_c$ as functions of $\mathrm{v}_0$.
            Both of the maps display the same sequence of intervals of $\mm{v}_0$\ where the outgoing variable ($H_0$\ or $N_c$)\ is smooth. Between these intervals there are gaps in which $H_0$\ and $N_c$ jump in an irregular way. Hence there is a correspondence between the number of crossings through the origin and the regular intervals for  $H_0$. 
       \item [-]  An analogous analysis is performed  with the driver $f_1$. The patterns of the scattering functions are qualitatively similar to the previous ones.   
       \item [-]  We initialize at $t=0$\ an ensemble of scattering orbits and we calculate the number $N(t)$\ of orbits that are still inside the interaction region at the time $t$. We obtain an algebraic decay law.
       \item [-]  For the same ensemble the distribution of zeros $N(n)$ is determined.\ For the considered parameters of the simulation we again obtain a power law.
       \item [-]  We fix the initial conditions and we calculate $H_0$\ and $N_c$\ as functions of the parameter $e_0$. The resultant plots show the same structures described for the previous scattering functions          
      \end{itemize}

  {\bf Note}

%     Numerical analysis is provided as for cases with a resonance occurrence as for other without any resonance; the latter is got whenever the external force frequency is larger than the 
%     inverse of the period of each bound orbit of the free oscillator.\\  
    %We arranged the parameter this way in order to avoid an effect which should not be pertinent to the item in hand. 
      A resonance may be a rich source of highly complex motions (see Ref.~\onlinecite{Haller1999}), as chaotic patterns and irreversibility in the transfer of energy between different oscillatory states. It takes place only if the system stays in one of those particular regions of the phase space in which the frequencies of some angular variables become nearly commensurate. 
      We guess that the irregular motions of the systems we consider are due to a mechanism distinct from and  more general than a resonance.
      Actually we avoid the resonance condition,\ since the driver frequency $\nu$ is larger than the highest 'effective' frequency of the free oscillator.

      \section{Numerical results.} 
   We use the  Runge-Kutta algorithm of fourth order to integrate the Hamilton's equations (\ref{def:hamiltonian}).  
%and we compute the Lyapunov's exponents by the technique that is provided in (\cite{Benettin}). \\
      In order to check our code, we solve the rigid spheres model and we compare with the same simulations in Ref.~\onlinecite{Kundu-Bauer}. 
      \subsection{Scattering functions}
% COMMENT
%\comment{ \begin{enumerate}
%\item La definizione delle variabili asintotiche. Se si usa il tempo lo spazio puo' essere considerato compatto per la periodicita' della forza esterna 
%\item Il fatto che ci sia un inviluppo \`{e} equivalente alla decrescenza a zero di un potenziale sinusoidale nel tempo? Pensa alla combinazione $\tau=t-(x/p)$
%\item it can be observed that the relevant qualitative characteristics of residual energy vs $e_{0}$\ graph are not affected if the parameters n, m of the external force are modified (let's compare among the figures\ 
%%%%%%(\ref{fig:finite_duration--final_energy_vs_e0--055-1.eps}), (\ref{fig:envse0_n100-m340_xo00_055.eps})) and (\ref{fig:envse0_n100-200_055})
%; the form close to $e_{0}=0$\ depends on the duration of switching on phase (given by n).
%\end{enumerate}}

      \begin{figure}[ht]
        \begin{center}
                   \includegraphics[width=0.6\textwidth]{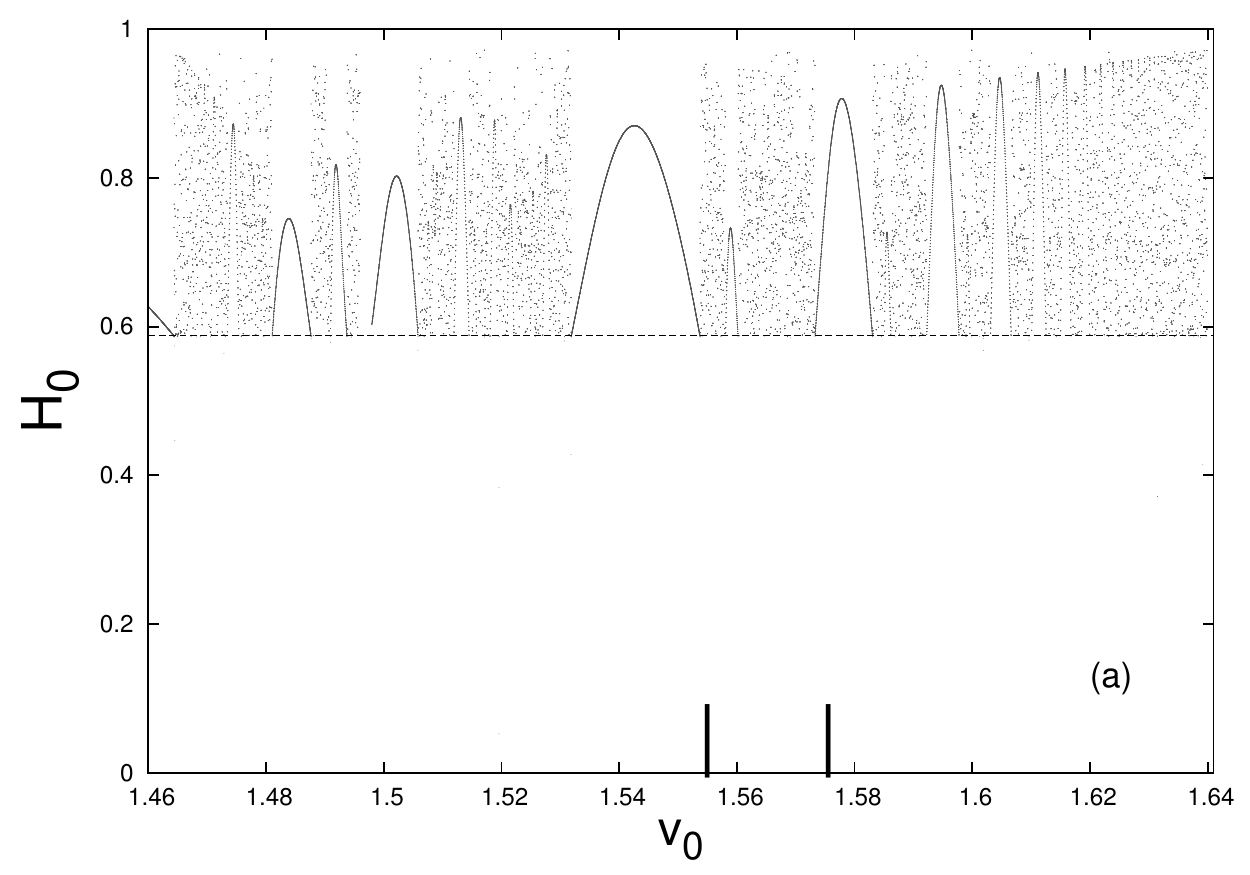}
                   \includegraphics[width=0.6\textwidth]{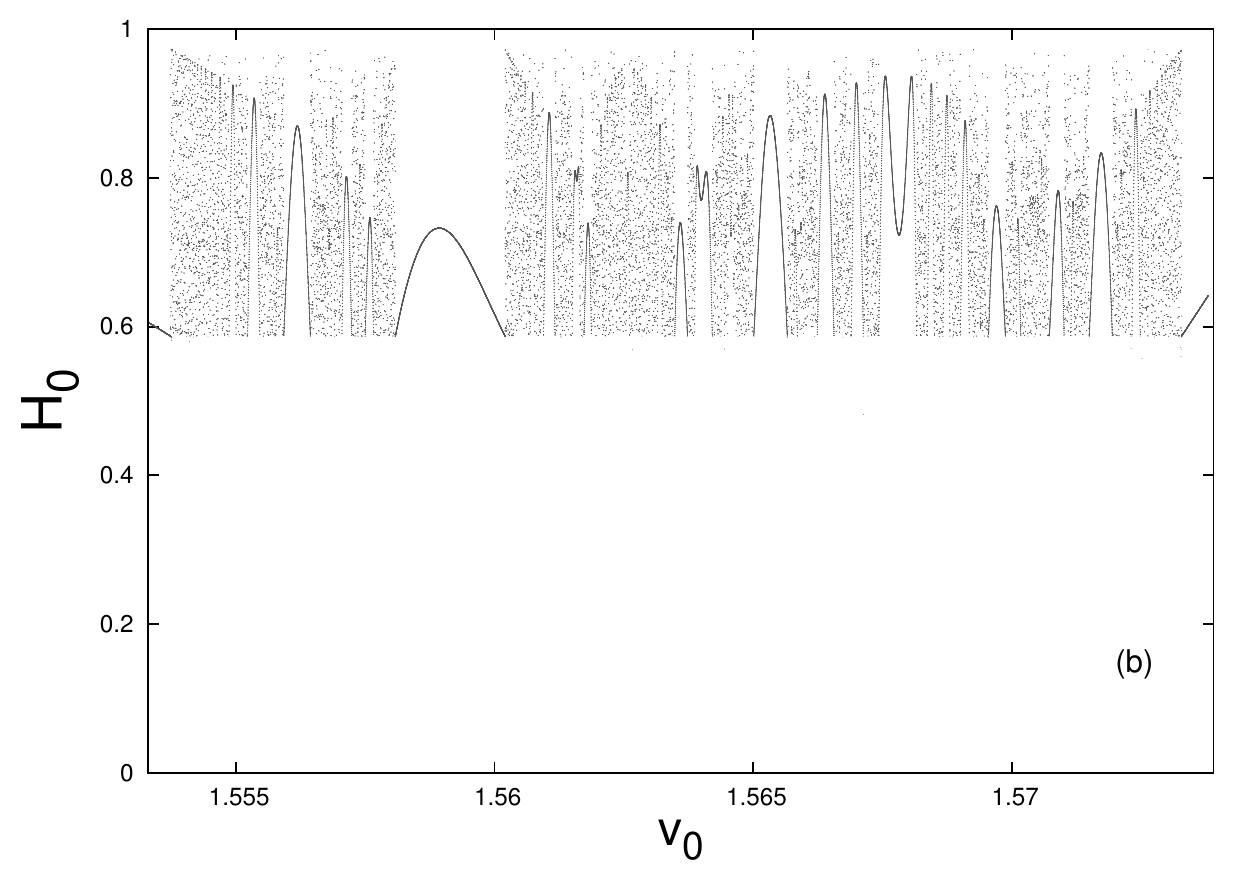}
          \caption{The map of the outgoing energy $H_0$ vs the initial velocity $\mm{v}_0$ for $x(0)=0$ with $e_0=1$\ and $\nu=0.8$. Plot $(b)$ is a magnification of the marked interval of plot 
          $(a)$.\label{fig:scattering-function}}
        \end{center}
      \end{figure}
%      Riguardo al modello occorre ancora parlare
%      \begin{itemize} 
%      \item [-] del fatto che l'ampiezza della forza esterna varia su un intervallo grande, da  valori piccoli, in modo che il sistema possa ancora essere considerato quasi-integrabile fino a valori anche di un ordine 
%      pi\'u grandi dei precedenti.
%      \end{itemize}
      A scatterig function is a map connecting some variables that refer to states before the interaction to some other concerning states after it.
      Our choice of the variables of the mapping is partially different from that in Refs.~\onlinecite{Beeker-Eckelt} and~\onlinecite{Eckelt-Zienicke}.

     When $f=f_1$\  we obtain the map    
     \[
        S^1_{(x_0,e_0)}:\ \ (x(0)=0,\ \mm{v}(0)=\mm{v}_0)\ \,\mapsto\ \,H_0(p(t^\ast),\,x(t^\ast))
    \]
    $t^{\ast}>\,n\,\pi/\nu$\,, $n=340$, i.e.  it links the initial conditions ($0$,\,$\mm{v}_0$)\ of a trajectory to the energy $H_0$ of the trajectory at a time $t^{\ast}$\ after the driver $f_1$\ is switched off. The driver amplitude $e_0$ is treated as a parameter. Analogous maps are obtained as functions of $x_0$\ for a given $\mm{v}_0$.
      When $f=f_2$\ (periodic case) we consider the map 
    \[
        S^2_{(x_0,e_0)}:\ \ (x(0)=0,\ \mm{v}(0)=\mm{v}_0)\ \,\mapsto\ \,H_0(p(t_k),\,x(t_k))
    \]
     $\nu\,t_k\ =\,2\,k\,\pi$\,,\ $k=1,\ 2\,,\ \ldots $.     
     It is a map from one value to many ones which  connects the initial conditions  ($0$,\,$\mm{v}_0$)\ to the energies $H_0$'s\ taken at times $t_k$\ such that $t_{k+1}-t_k$\ is one driver period.

In all the cases  the dependence of $H_0$\ on $\mm{v}_0$\ alternates between regular intervals, where $S^i_{(x_0,e_0)}$\,, $i=1,\,2$, shows a smooth behavior, and irregular intervals, where  $S^i_{(x_0,e_0)}$\ is much more sensitive. In the plot $(a)$ of fig.~\ref{fig:scattering-function} we present the results for the $\mm{v}_0$\ interval $[1.46,\,1.64]$.
Plot $(b)$ of fig.~\ref{fig:scattering-function} is an expansion of the irregular interval marked in the plot (a) ($\mm{v}_0$\ varies in $[1.5533:1.5739]$).
%%%    We note that the pattern is qualitatively similar to the complete one; the behavior of an irregular gap in $(a)$\ of (fig.-\ref{fig:scattering-function}) dissolves into a sequence regular subintervals separated by 
%%%    irregular gaps.\\ 
%%%    Whenever we continue in this way we find an intertwined structure of regular and irregular intervals. 
%%%    From our numerical inspection we are driven to believe that $S^1_{(x_0,\,e_0)}$\ should possess  fractal features; the scattering function is continuous for all $\mm{v}_0$\ except a set $\mathcal{C}$\ of singularities.\ 
%%%    $\mathcal{C}$ is the same as the set of points at the boundaries between each couple of contiguous regular intervals.

%%%    We guess the measure of $\mathcal{C}$\ should be zero. In fact $\mathcal{C}$\ appears to be determined in analogy to the hierarchical procedure of a Cantor set, by giving level by level the endpoints of cutout intervals.
      \begin{figure}[ht]
        \begin{center}
          \includegraphics[width=0.45\textwidth]{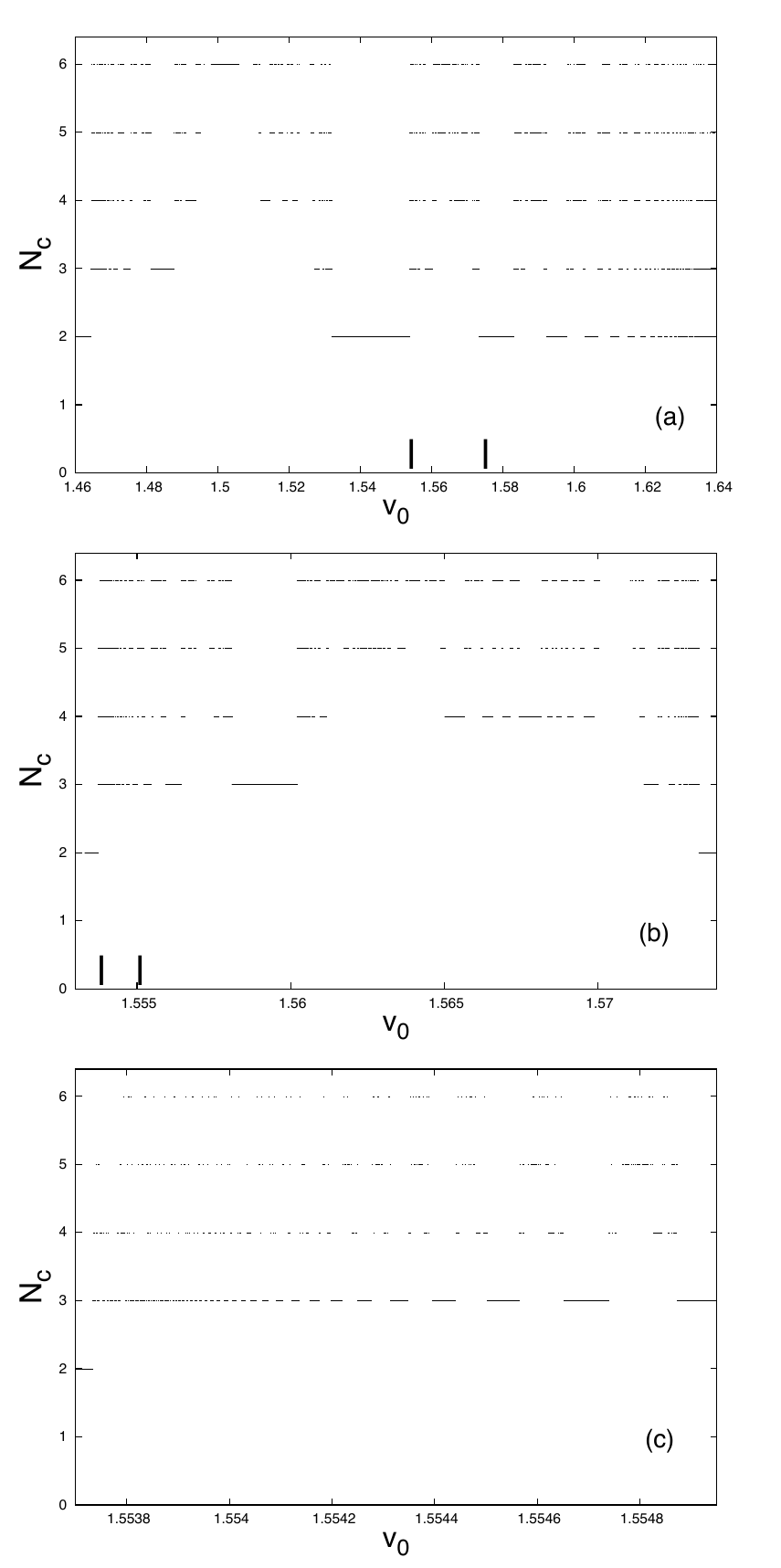}
          \caption{The number of zeros $N_c$ of a trajectory with initial velocity $\mm{v}_0$ ($x(0)=0$); the marked interval of $(a)$\ is magnified in $(b)$; the plot $(c)$\ shows a sequence of intervals with three 
           zeros that accumulates to a two zeros interval as $\mm{v}_0$\ approaches the border of the two zeros interval. \label{fig:crossing-zeros} }
        \end{center}
      \end{figure}
%     For each scattering orbit there exists at least one time $t^0>0$\ such that $x(t^0)=0$: in fact both of $V_i$'s are attracting and then a scattering interaction needs one passing through $x=0$. 
An orbit may be marked by the number $N_c$, the number of crossings of $x=0$. Another scattering map we obtain is $N_c$\ as a function of $\mm{v}_0$; it is displayed  in 
the part $(a)$ of fig.~\ref{fig:crossing-zeros}. We note the following features: 
      \begin{itemize}
        \item [-]\ $N_c$\ is constant in each interval where the scattering map  is regular; $N_c$\ and the scattering map are singular at the same points (compare (a) of 
        fig.~\ref{fig:scattering-function} to (a) of fig.~\ref{fig:crossing-zeros});     
        \item [-]\ each regular interval between two regular ones with the same $N_c$\ has a crossing number greater than $N_c$; 
        \item [-] in the region between two regular intervals with the same $N_c$, a sequence of $N_c+1$ regular intervals accumulates as $\mm{v}_0$ gets next to the boundaries of the region (see  (b) and (c)\ of 
         fig.~\ref{fig:crossing-zeros}).   
      \end{itemize}
     \begin{figure}[ht!]
       \begin{center}
         \includegraphics[width=0.45\textwidth]{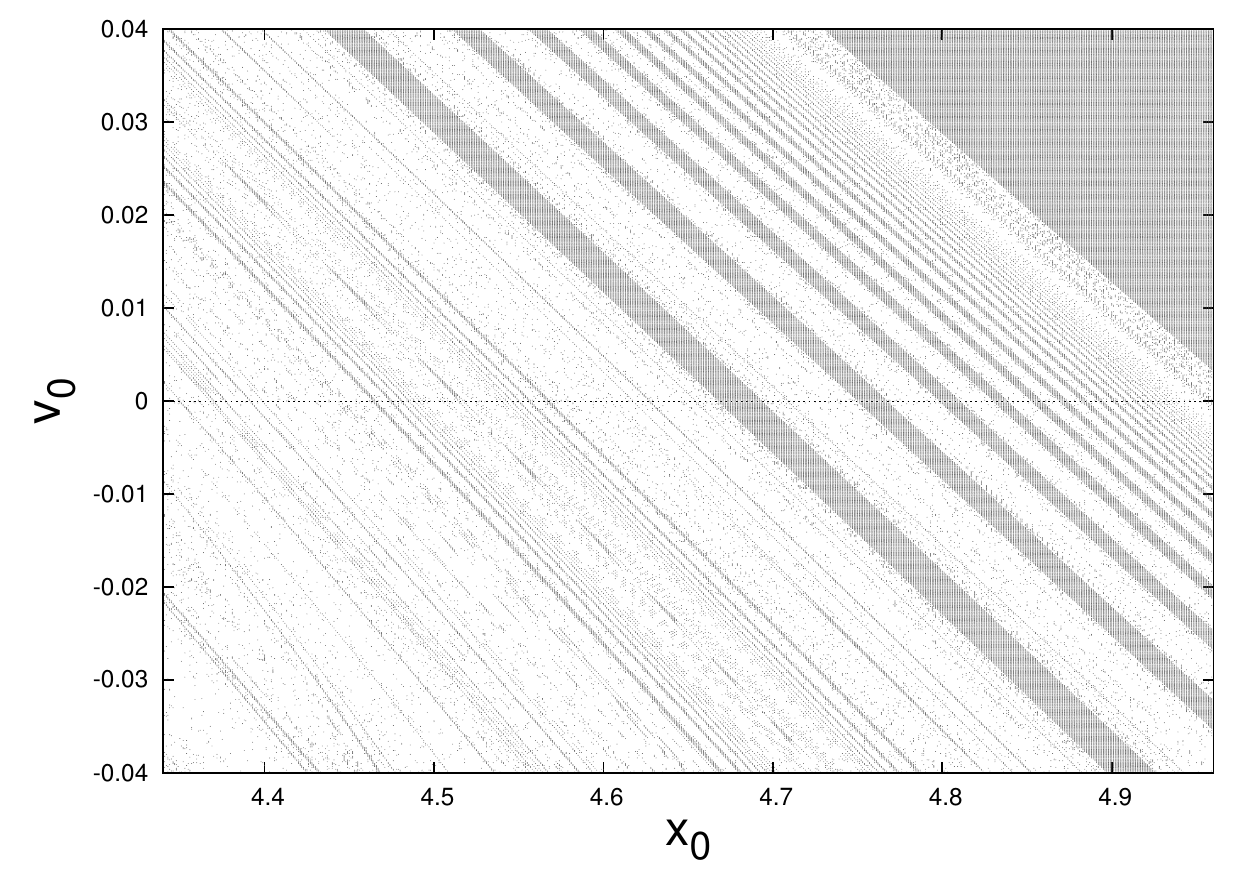}
         \caption{Distribution of initial conditions of nearest trajectories with the same number of crossings of the origin. We choose an ensemble of orbits with an uniform distribution of initial conditions and we mark an initial condition whenever its orbit experiences the same $N_c$ as the orbits from the closest initial conditions.}\label{fig:fractal}
       \end{center}
     \end{figure}   
    \begin{figure}[ht!]
       \begin{narrow}{-1cm}{0cm}
       \begin{centering}
         \includegraphics[width=0.5\textwidth]{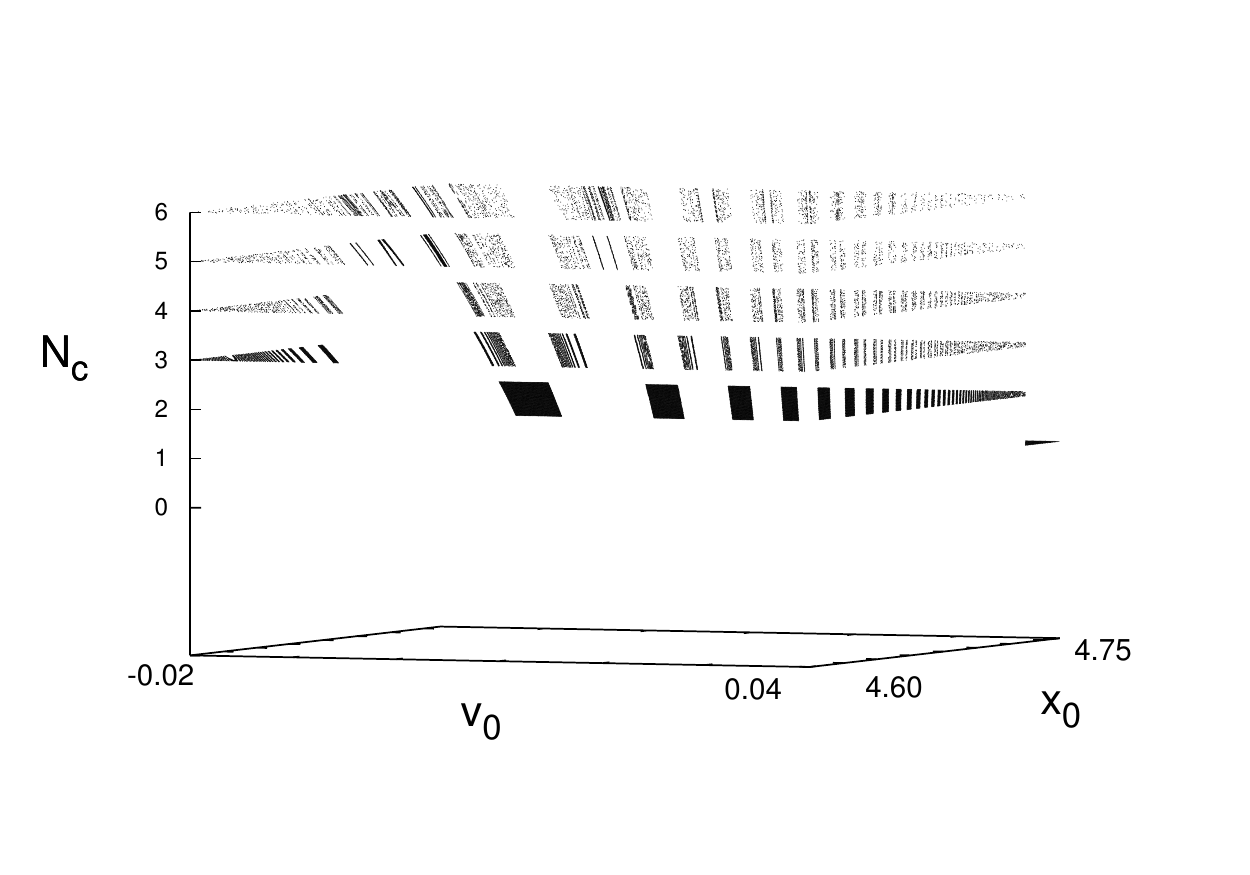}
         \caption{The plot shows $N_c$ versus  the initial conditions in the same numerical experiment as in fig. \ref{fig:fractal}. Note the stripes arrangement of the structure and the hierarchy of stripes according to what described above. The $N_c$ stripes accumulate going close to a $N_c-1$ stripe.}\label{fig:N_c--vs--x_0-v_0.eps}
       \end{centering}
       \end{narrow}
     \end{figure}
         \begin{figure}[ht]
        \begin{center}
          \includegraphics[width=0.5\textwidth]{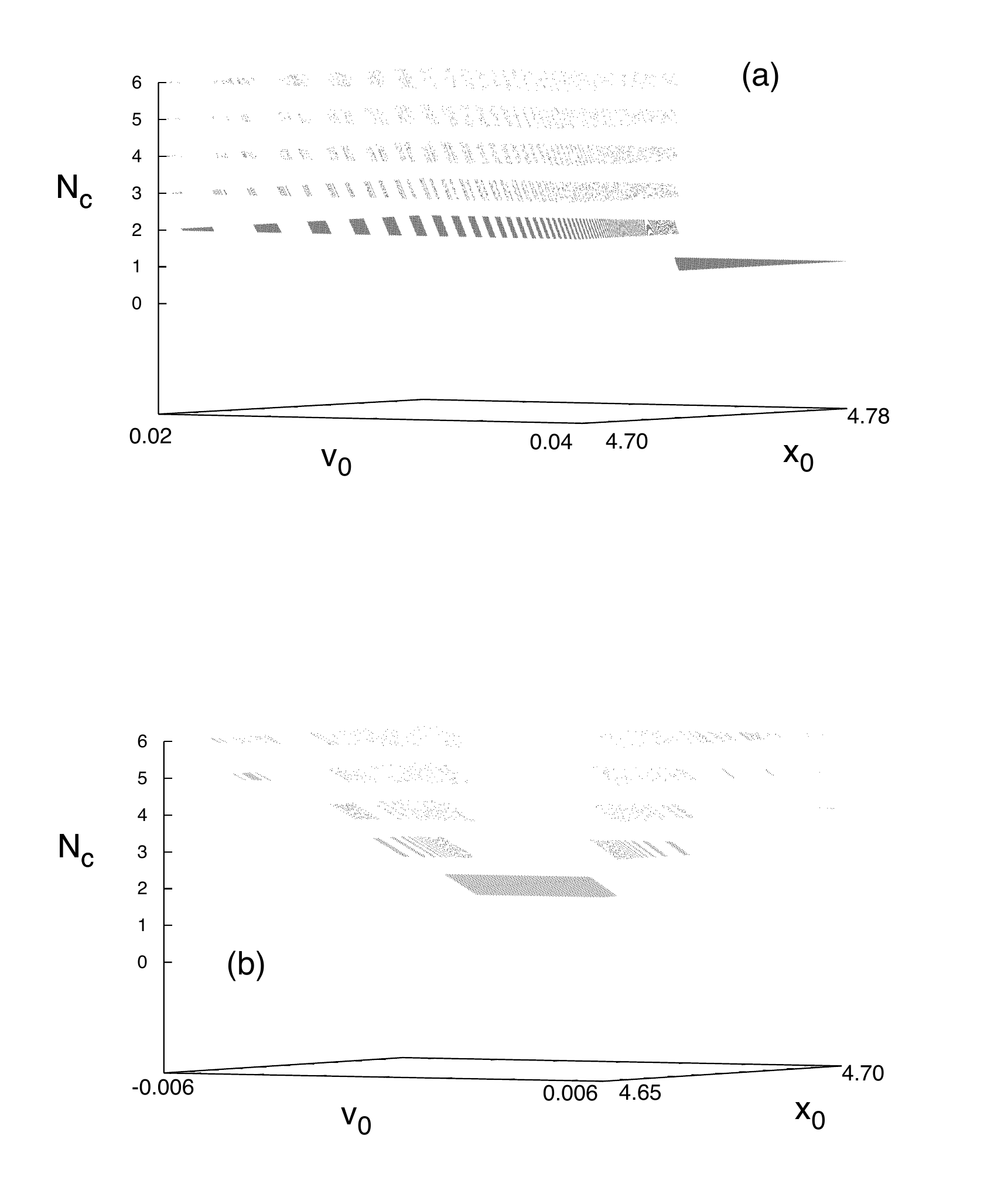}
          \caption{There are shown some details of fig. \ref{fig:N_c--vs--x_0-v_0.eps}. In $(a)$ there is a sequence of stripes $N_c=2$\ approaching the domain $N_c=1$. $(b)$\ is an example of the structure with a 
           sequence of $N_c+1$ stripes accumulating against a $N_c$\ domain.   
          \label{fig:N_c--vs--x_0-v_0-1.eps} }
        \end{center}
      \end{figure}   
      These rules define the 'hierarchy' of the structure of the scattering function and of the delay time function.

A trajectory belonging to a regular interval spends a finite time in the interaction region since, we know, it experiences a finite number of zeros.  When the trajectory leaves out the interaction region, only the driver has an effect on the particle and therefore the values of\ $H_0$\ at times $t_k$,\ taken at the end of each driver period (see the definition of $S^2_{x_0,\,e_0}$),\ are almost constant, i.e. for $\mm{v}_0$\ in a regular interval the map $S^2_{x_0,\,e_0}$\ is characterized by a 'fixed' point (see fig.~\ref{fig:scattering-function}).

The horizontal line in fig.~\ref{fig:scattering-function} - $(a)$\ displays the energy $h'$\ of the unperturbed escape orbit ($0.588 $ in our cases). From the plot we may state that  if $\mm{v}_0$\ is in a regular interval then $H_0\ge h'$\ for all times large enough. An orbit belonging to a regular region is called hyperbolic and parabolic  when $H_0 >h'$\ and $H_0=h'$, respectively. 

We note that a magnification of an irregular interval between two regular intervals with $N_c$ zeros gives a pattern qualitatively similar to the complete one; the irregular gap in $(a)$\ dissolves into a sequence of regular subintervals, with at least $N_c+1$\ zeros, separated by irregular gaps. 
    Whenever we continue in this way we find an intertwined structure of regular and irregular intervals; at each step the distribution of regular intervals, marked by the number of zeros, meets the three rules listed 
    above.
  
    From our numerical inspection we are driven to argue that $S^2_{(x_0,\,e_0)}$\ should possess  fractal features; the scattering function is continuous for all $\mm{v}_0$\ except a set $\mathcal{C}$\ of singularities.\ 
   $\mathcal{C}$\ is the set of points after removing all the regular intervals with any number $N_c$\ of zeros. 
    We guess the measure of $\mathcal{C}$\ should be zero. In fact $\mathcal{C}$\ appears to be determined in analogy to the hierarchical procedure of a Cantor set, by giving level by level the endpoints of cut out intervals.

    Fig. \ref{fig:fractal} illustrates the distribution of the nearest trajectories which cross the origin the same number of times. Starting from a uniformly distributed ensemble of initial conditions, we mark a pair
   $(x_0,\,\mm{v}_0)$ of initial conditions  
   only if its trajectory experiences the same number of zeros as the trajectories corresponding to the closest pairs. The plot shows a pattern organized as a sequence of stripes accumulating to a domain corresponding to parabolic and hyperbolic orbits with only one zero, i.e. 
    to the set of initial conditions of orbits that escape after crossing once the origin. Each stripe in the figure is marked by a determined $N_c$.

    Fig.~\ref{fig:N_c--vs--x_0-v_0.eps} displays the distribution of the number of zeros $N_c$,\ with $1\le N_c \le 6$, extracted from the fig.~\ref{fig:fractal}. We observe the hierarchical structure, ordered with respect to $N_c$ and 
    described above. In particular we point out that the pattern is arranged in sequences of stripes that accumulate at the border of the domain with $N_c=1$ (see $(a)$ in fig.~\ref{fig:N_c--vs--x_0-v_0-1.eps}). 
%i.e. the set of initial conditions of orbits that escape after crossings once the origin (parabolic, hyperbolic orbits for $t\rightarrow +\infty$). 
 In the hierarchy of the structure the level $N_c+1$\ is a 
    sequence of stripes approaching the stripe $N_c$ (see $(b)$\ in fig. \ref{fig:N_c--vs--x_0-v_0-1.eps}) as well. 

A similar  hierarchy is observed  by Eckelt at al. \cite{Eckelt-Zienicke} for a class of oscillators characterized by attractive potentials.   In order to explain the properties of the hierarchy, they use a map (return map) introduced by Alekseev \cite{alekseev}.
    We guess that the mechanism  acting in our problem should be treatable in the same way by a suitable return map: we will come back to it in the next section.
%\comment{In particolare notiamo la sottostruttura in strisce della striscia n che si accumnulano verso il bordo della striscia n-1. Noi introdurremo una mappa D (mappa del ritorno) che spiega questa 
%    complessita di strutture.\\ 
%    Noi non affermiamo di aver dimostrato numericamente the appearance of chaotic scattering, i.e of a Cantor set of singularities. However le proprieta della struttura che osserviamo sono le stesse che osservano
%    Eckelt Zenieke even if our potential is not a representative of the class of potential they study  .  They introduce   }.
%    A trajectory belonging to a regular region spends a finite time in the interaction regions since, we know, it experiences a finite number of zeros.  When the trajectory leaves out the interaction region nearly only the 
%    driver has an effect on the particle and therefore the value of\ $H_0$\ at time $t_k$,\ following with the driver period (see the definition of $S^1_{x_0,\,e_0}$),\ is almost constant; i.e., in 
%    (fig.-\ref{fig:scattering-function}),  for $\mm{v}_0$\ in a regular interval  the map $S^1_{x_0,\,e_0}$  is characterized by a 'fixed' point. \\ 
%    The horizontal line in $(a)$\ (fig.-\ref{fig:scattering-function}) displays the energy $h'$\ of the unperturbed escape orbit ($0.588$). From the plot we may state that  if $\mm{v}_0$\ is in a regular interval then 
%    $H_0\ge 0.588$. An orbit belonging to a regular region is called hyperbolic and parabolic  when, respectively, $H_0 >0.588$\ and $H_0=0.588$. 
    \begin{figure}[ht!]
       \begin{narrow}{-1cm}{0cm}
       \begin{centering}
         \includegraphics[width=0.4\textwidth]{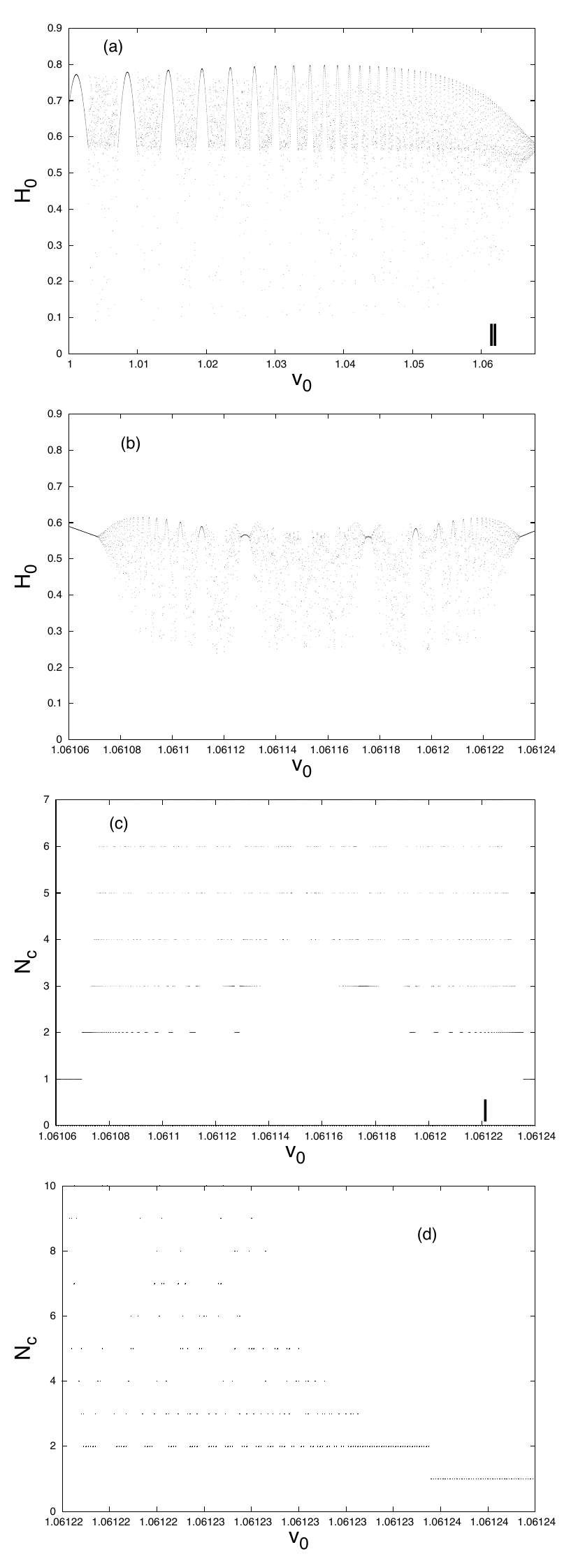}
         \caption{Scattering map and $N_c$ function of the oscillator (\ref{eq:oscillatore-lorentz}) with the driver (\ref{def:gtn}). Plot $(b)$\ is an enlargement of the marked interval in $(a)$.
           Plot $(d)$\ is a magnification of the marked interval in $(c)$. We note that the hierarchical structure is not fully developed.}\label{fig:duratafinita}
       \end{centering}
       \end{narrow}
     \end{figure}
    We also consider the non periodic case ($f=f_1$) and compute the map $S^1_{(x_0,\,e_0)}$ that is calculated when the driver is already switched off and the energy $H_0$\ is constant. 
%As above an orbit for $\mm{v}_0$\ in a  regular interval is called hyperbolic and parabolic when, respectively, $H_0 >0.588$\ and $H_0=0.588$.\\
    The plots in $(a)-(d)$ of fig.~\ref{fig:duratafinita} display the results; we set $e_0=1$\ and we choose $\nu=1$\ and $n=360$\ in (\ref{def:gtn}).   
    As above an orbit for $\mm{v}_0$\ in a  regular interval is called hyperbolic and parabolic when\ $H_0 >h'$\ and $H_0=h'$\ respectively. \\
    $S^1_{(x_0,\,e_0)}$ displays structures that seem qualitatively similar to $S^2_{(x_0,\,e_0)}$; but note that the hierarchical structure does hold only partially. Between two $N_c$\ intervals there are no  
    intervals with a number of crossings less than $N_c$. However some sets of orbits result to be lost: in particular they are  orbits in the vicinity of parabolically escaping ones (see $N_c \ge3$ orbits in 
    $(d)$ of fig.~\ref{fig:duratafinita}). These orbits have a very long return time and the number of their $x=0$\ crossings depends on the duration of the driver in a more considerable way.     
%    We may repeat the observations concerning the map $S^1_{x_0\,e_0}$,  without any changement.\\
    This result is expected. A driver of infinite duration is needed so that the map structure can display a complete hierarchical pattern; a fractal generating mechanism needs an infinite  
    iterative process.  

    \subsection{The delay time function and scaling properties.}
       We now focus on the relation between the time of permanence in the interaction region and the regularity of the motion. The time interval spent by the particle in the interaction region is called delay time.

       Given an ensemble of orbits (initial conditions) we determine the 'survival function' $N(t)$: it gives the  number of orbits that have a delay time larger than $t$, i.e. the number of orbits that experience at least one crossing through $x=0$ at $t_1\ge t$. 
      This definition is not exactly the same as that considered by some other authors
       \cite{Lai-Blumel-Ott-Grebogi-1992,Ding-Bountis-Ott,Beeker-Eckelt} and which needs a bound region around $x=0$ to be initially chosen and counts the number of orbits that still stay in the region at 
       $t$.   
       
\begin{figure}[!h]
         \begin{narrow}{-1cm}{0cm}
           \centering
           \includegraphics[width=0.5\textwidth]{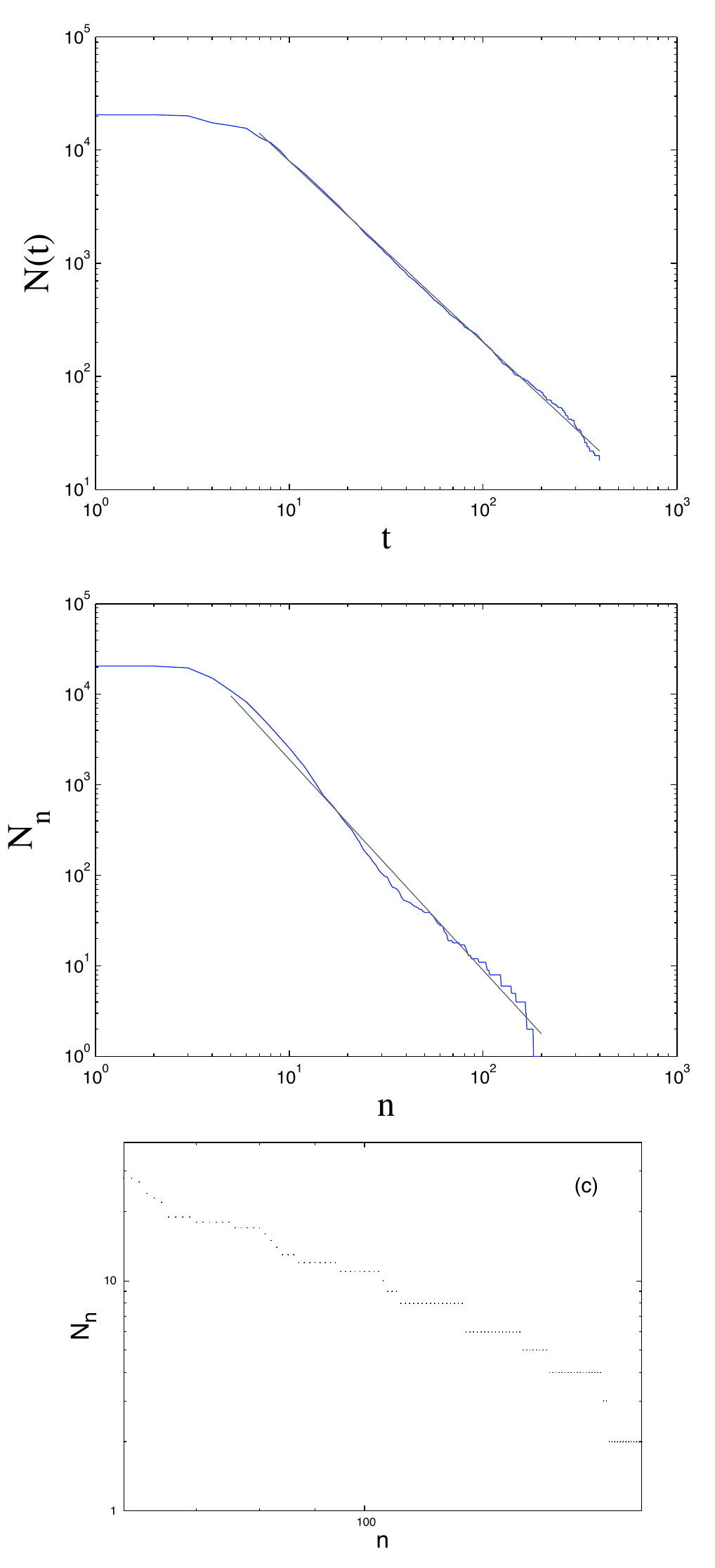}
           \caption{\label{fig:decay_function--e0_1-x0_0-omega_08-1_55} Log-log plots of $N(t)$ (plot $(a)$) and $N(n)$ (plot $(b)$)\ obtained from a sample of $2\ 10^4$\ orbits whose initial velocity is 
           uniformly chosen in (1.55,\ 1.57). Plot $(c)$\ is an enlargement of the tail of the plot $(b)$\ from the marked point; it shows a 'stair-case' behavior of the distribution $N(n)$.}    
         \end{narrow}
       \end{figure}

       In our case the set of initial conditions is defined by a uniform distribution of initial velocities in the interval $1.55\le \mm{v}_0 \le 1.64$\ and $x(0)=0$. The line of initial conditions intersects the stable manifolds of the 
       invariant set transversally. 
A scaling law is extracted by plotting $N(t)$\ vs $t$ in a log-log diagram shown in fig.~\ref{fig:decay_function--e0_1-x0_0-omega_08-1_55}.  %a log-log plot displays that $N(t)$ appears to have a long time algebraic decay: $N(t) \sim t^{-z}$\ with $z=1.6$.\\
% It exists a time $\overline{t}\simeq 100$\ such that in both intervals $(0,\,\overline{t})$\ and $(\overline{t},\,200)$\ 
The plot is fitted by a straight line, i.e. $N(t)$\ 
 displays an algebraic behavior $N(t)\sim t^z$; the exponent is $z=1.60\pm 0.01 $. 
%while the exponent over the remaining interval is $z_2=2.38\pm 0.04$. 

The power law in the long time tails of the delay time statistics is related to the non-hyperbolicity of the dynamics. We expect a behavior like $t^{-3/2}$\ when KAM islands influence the scattering process  (see Refs.~\onlinecite{Karney},~\onlinecite{Chirikov-Shepelyanski}) and $t^{-2}$ when parabolic surfaces influence the scattering process, but KAM islands do not.
%%       For $t$\ large enough ($t > 100$) $N(t)$\ appears to be constant; this should be correlated to the regular intervals of the scattering function. 
%\comment{aggiungere osservazione}
%(\comment{\textsl{questa affermazione deve essere verificata }).}\\    
%       $N(t)$\ decays according to an algebraic law  where it\ changes;\ 
%\comment{the decay exponents are equal or not?} \textsl{occorrono delle simulazioni con una maggiore risoluzione sulle condizioni iniziali}.

       Another tool, that can be helpful to extract a statistical measure of the irregular dynamics in a scattering problem, is the distribution of the number of zeros.
        We consider the same ensemble of orbits as above; we determine the number $N(n)$\ of orbits of the ensemble that perform at least $n$\ crossings through\ $x=0$.
        We obtain again an algebraic law: $N(n)\sim n^{-z}$ for $z= 2.33\pm 0.05$ (see fig.~\ref{fig:decay_function--e0_1-x0_0-omega_08-1_55} - (b)).
        The power-law holds for $5\le n\le 60$.\ For $n > 60$\ $N(n)$\ consists of a 'staircase' structure 
        (see fig.~\ref{fig:decay_function--e0_1-x0_0-omega_08-1_55} - $(c)$) where the jumps take place at an irregular sequence of points.
%        We note that  for $t$\ large enough ($t > 100$) the delay-time function $N(t)$\ has the same feature. It is likely that these facts are not independent. 
         In Ref.~\onlinecite{Beeker-Eckelt}\ it is obtained that $N(n)$ follows an exponential decay. The discrepancy is a consequence of a different choice of the parameters. We will come back to this point later.

\section{The return map.}
In this section we deal with a mapping that allows to reduce the analysis of the complete dynamical problem (\ref{def:hamiltonian}) to that of the states that
either will still undergo at least a zero in the future or already underwent at least a zero in the past. Using this approach we can pick out an arrangement of the states that
provides an explanation of the scattering map structure. 
%In this section we show that a horseshoe is formed in systems (\ref{def:hamiltonian}). 
%Similar results are found in (\onlinecite{alekseev}) (see also (\onlinecite{Eckelt-Zienicke})) where irregular properties of the dynamics in periodically oscillating potential wells are investigated.
%Let us note that the construction of the horseshoe is local, only in the neighborhoods   of some points in the phase space.\\ 
 In this section we deal with the periodic driver $f_2$.   

The considered map, called return map, is introduced  in the Refs.~\onlinecite{alekseev} and~\onlinecite{Eckelt-Zienicke}, where the authors investigate the irregular properties of the dynamics in periodically oscillating potential wells; they show that the map locally possesses the properties of the horseshoe. 
%It produces an invariant set that is relevant to explain the features of the scattering function. 
In our case we cannot conclude about the hyperbolicity of the invariant set because a better behavior of the potential $U(x,\,t)$,\ for $|x|\rightarrow +\infty$,\ is needed. 
%However on the basis of some qualitative considerations 
%the procedure to determine the invariant set of the map allows us to introduce some sets of the phase space whose structure may be straightforwardly compared with our numerical results on the scattering functions. 

     We define the return map $D$ on the Poincar\'e section $x=0$, that we denote by $\Gamma$.
     Consider a motion in which the particle crosses $x=0$\ at the instant $t$,\ with momentum\ $p$.\ The conditions $(x=0,\,p,\,t)$ \ uniquely determine the solution $x(t';\,p,\,t)$\ of the problem (\ref{def:hamiltonian}). Whenever it exists consider the minimum time $t_1$,\ $t_1>t$, such that $x(t_1;\,p,\,t)=0$. The return map is defined by 
%     Let us note that
%     \be
%       x(t;\,p_k,\,t_k )\,=\,x(-t;\,-p_k,\,t_k)\,=\,x(\nu\,t\,+\,2\,\pi ,\,p_k ,\,\nu\,t_k\,+\,2\,\pi)
%     \ee
%     since $H(x,\,p,\,t)=H(x,\,-p,\,-t)=H(x,\,p,\,\nu\,t\,+\,2\,\pi )$. It follows that 
 %    In particular, the set\ of zeros\  $\{x(t\,;\ p_k,\,t_k)=0\}$\, is uniquely fixed by $(p_k,\,t_k)$,\ 
     \be
        D\ :\ \ (p,\,t)\ \longrightarrow\ (p_1,\,t_1)
     \ee   
      where $p_{1}$\ is the momentum at $t_{1}$.

\begin{figure}[!h]
         \begin{narrow}{-0.6cm}{-0.cm}
           \centering
           \includegraphics[width=0.42\textwidth]{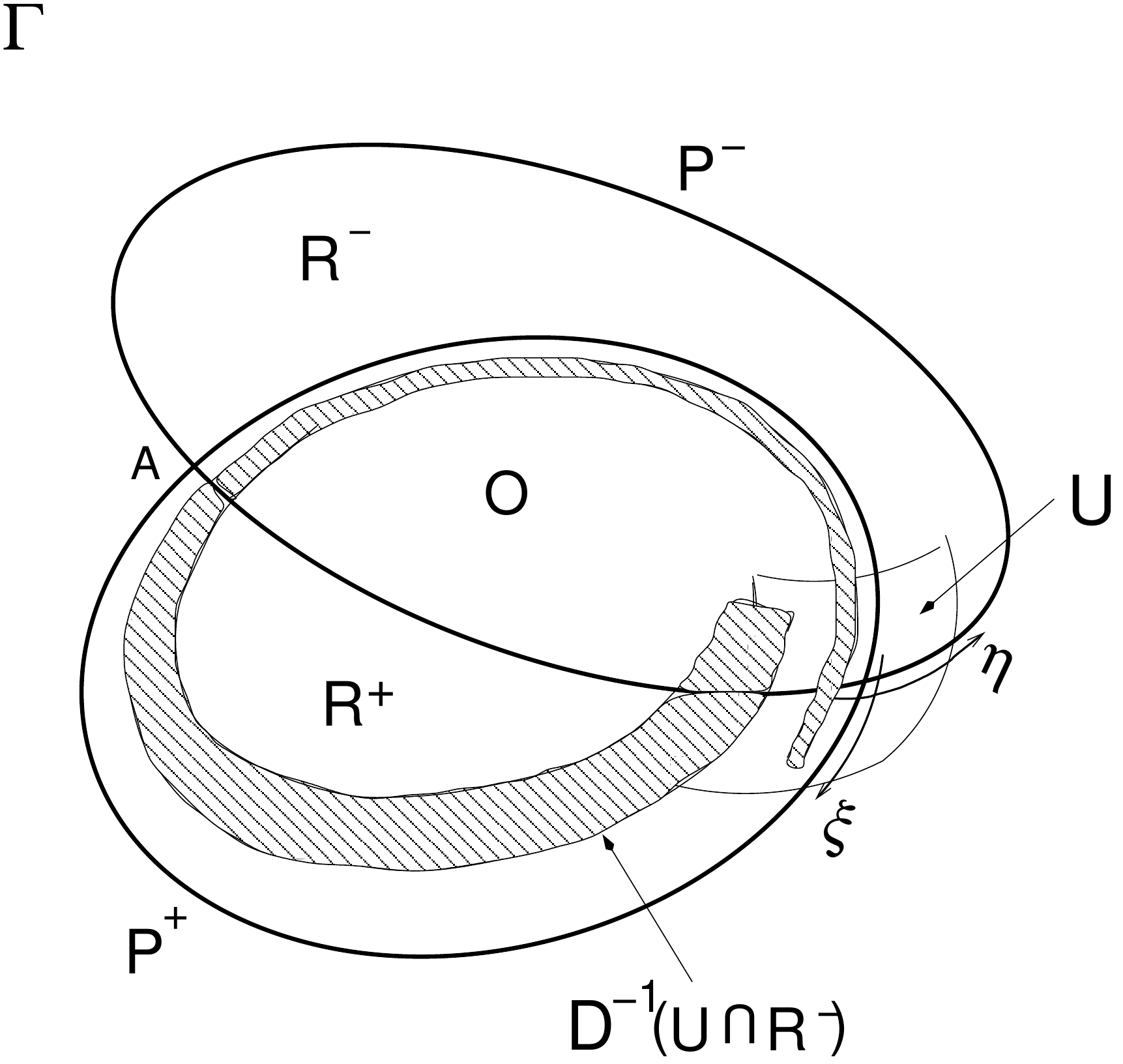}
           \caption{\label{fig:return-map-1} The panel shows the sets $R^+$\ and $R^-$ and their boundaries\ $P^+$ and $P^-$ respectively. $U$ is the neighborhood of a regular point where a system of local coordinates
              $(\xi,\,\eta )$\ is introduced. The pre-image of $U\cap R^-$\ under the return map is shown.}
         \end{narrow}
       \end{figure}

      Note that
     \be
       x(t';\,p,\,t )\,=\,x(-t';\,-p,\,t)\,=\,x(t'\,+\,2\,\pi/\nu ;\,p ,\,t\,+\,2\,\pi/\nu)
     \ee
     since $H(p,\,x,\,t)=H(-p,\,x,\,-t)=H(p,\,x,\,\nu\,t\,+\,2\,\pi )$.\ It follows that the zeros of the solutions of problem (\ref{def:hamiltonian}) can be marked by the 'polar' coordinates  $p \ge 0$\ 
     and\ $\tau\equiv\nu\,t$.
      We denote by \,$R^+\subset \Gamma$\ \,the domain of $D$ (see fig.~\ref{fig:return-map-1}).  
%and  $R^-$\ the image\  $\{(p_{k+1},\,\tau_{k+1})=D(p_k ,\,\tau_k);\ (p_k ,\,\tau_k)\in R^+\}$\ of $R^+$.
%     We note that $\partial_xU(x,\,t)\,=\,0$\ for all $t$; from this it follows\ $(0,\,\tau)\in R^+$.  $R^+$\ is a nonempty set; as well $R^+$ is open (see~\onlinecite{alekseev}, theorem 1).
     Let us consider $(p,\,\tau)\in R^+$\ and $(p',\,\tau')=D(p,\,\tau)$. There exists the maximum (whenever $x \ge 0$)\ or minimum (whenever\ $x\le 0$) $X^+(p,\,\tau)$\ of $x(t;\,p,\,\tau)$\ in the interval 
     $(\tau,\,\tau')$.  
%and $x\,>\,0$\ for $\tau_k <t <\tau_{k+1}$;  
%%      we consider the next turning point \ $X^+(p,\,\tau)$   
%=\sup_{\tau_k <\nu\,t <\tau_{k+1}}x(t;\,p_k,\,\tau_k)$\,, 
%%      of the actual zero $(p,\,\tau)$.

      In our case the energy is not conserved, but one can, all the same, define an energy-like function. For all $(p,\,\tau)$\ in $R^+$ it is given
%      It is defined the function 
         \be\label{def:energy-h}
       h^+(p,\,\tau)=
%      \left \{ 
%         \begin{array}{lc}    
             \mathcal{V}(X^+(p,\,\tau))  
%                        &      |X^+| < +\infty \\
%%%             \dot{x}(+\infty ;\,p,\,\tau )^2/2\,+\,h' & |X^+|= +\infty
%         \end{array}  
%     \right .
     \ee
%     $h^+(p_k,\,\tau_k)=V(X(p_k,\,\tau_k))$, and 
     where \ 
%$h'=V(+\infty)$ and $V=V_i$, $i=1,\,2$ 
      $\mathcal{V}$\ is the time average 
      \[
         \mathcal{V}(x)\,=\,1/(2\,\pi)\int_0^{2\,\pi}d\tau'\ U(x,\,\tau')     
      \]
      It is straightforward to show that $\mathcal{V}=V=V_i$,\ $i=1,\,2$. The function $h^+$ a composition of continuous applications; hence $h^+$\ is a continuous function in $R^+$.
      We can extend by continuity the definition of $h^+$ at the boundary $P^+$\ of $R^+$; $P^+$ consists of points $(\tilde{p},\,\tilde{\tau})$\ such that if $(p,\,\tau)\in R^+$\ approaches $(\tilde{p},\,\tilde{\tau})$\ then 
      $|X^+(p,\,\tau)|\rightarrow +\infty$\ and    
      \[
         h^+(\tilde{p},\,\tilde{\tau})\,=\,\lim_{(p,\,\tau)\rightarrow\,(\tilde{p},\,\tilde{\tau})}h^+(p,\,\tau)\,=\,V(+\,\infty)
      \]      
      A point $(p,\,\tau)$ in the set $H^+=\Gamma\setminus (R^+\cup P^+)$\ is connected to an orbit $x(t;\,p,\,\tau)$\ whose $X^+(p,\,\tau)$ is infinite; it follows that $\dot{x}(t;\,p,\,\tau)^2/2+V(x(t;\,p,\,\tau))\ge V(+\infty)$\ for all $t$\ large enough. The orbit $x(t;\,p,\,\tau)$\ is called hyperbolic\ for $t\rightarrow +\infty$. 
%%      $h^+$\ enables us to classify the solutions just as this is done with a constant energy; referring to the above definitions a trajectory $x(t,\,p,\,\tau)$\ is parabolic and hyperbolic when 
%%      $h^+(p,\,\tau)=h'$\ and\ $h^+(p,\,\tau) > h'$\ respectively.\\   
%     The function $h^+$ is defined by a composition of continuous applications; hence $h^+$\ is a continuous function in $\Gamma$. 
%%     We consider the set $P^+=\{(p_k,\,\tau_k)\,;\ h^+(p_k,\,\tau_k)=h'\}$\ and $H^+=\Gamma\setminus (R^+\cup P^+)$;  $P^+$\, and $H^+$\ are non-empty. 
%and $x(t;\,p\,\tau)$\ is parabolic and hyperbolic when $(p,\,\tau)\in\Pi^+$ and, respectively,  $(p,\,\tau)\in\ H^+$ (see~\onlinecite{alekseev}, theorem 1).

     We may introduce analogous quantities following the solutions backwards in time.
     We focus on the set of all zeros $(p,\,\tau)\in\Gamma$ for which there exists $x(t';\,p,\,\tau)=0$\ at $t'< \tau/\nu$, i.e. the particle returns to the origin if we follow the solution back to the past. We denote this 
     set by\ $R^-$ (see fig.~\ref{fig:return-map-1}); it is the imagine of $R^+$\ under the return map. 
     If $(p,\,\tau)\in R^-$\  then it exists a finite last turning point $X^-(p,\,\tau)$.\ The function\ $h^-(p,\,\tau)$\ is defined in $R^-$,\ analogously to (\ref{def:energy-h}); now substitute $X^-$\ for\ $X^+$. $h^-$\ is continuous in $R^-$ and can be extended to the boundary\ $P^-$\ of $R^-$.
%     We consider the set $P^-=\{(p,\,\tau)\ ;\ h^-(p,\,\tau)=h'\}$\ where $h'=V(-\infty)$ ($V(x)=V(-x)$). 
We introduce\ $H^-=\Gamma\setminus (R^-\cup P^-)$, as well. The point\ $(p,\,\tau)$ is\ in $H^-$\ whenever $X^-(p,\,\tau)$\ is infinite; the solution $x(t,\,p,\,\tau)$\ is called hyperbolic for $t\rightarrow\,-\infty$.   
%$P^-$\ and $H^-$\ are non-empty and, respectively, closed and open (see~\onlinecite{alekseev}, theorem 1).\\    
%     By the definitions we state that $X^-\circ D=X^+$. It follows that\ $h^-\circ D= h^+$; hence\ $R^-=D(R^+)$\ and\ $P^-=D(P^+)$. 
%     We call $x(t;\,p,\,\tau)$ oscillatory if $(p,\,\tau)\in (R^+\cap R^-)$\ and\ $D^{n}(p,\,\tau)\in (R^-\cap R^+)$\ for all\ $n\in\mathbb{Z}$.

%     Let $\tau'$\ and $\tau''$\ be, respectively, the time of the next following zero and the time of the last previous zero of $x(t;\,p\,,\tau)$. 
%     The following property plays an important role.  

     If  $(p,\,\tau)$\ is in $R^+$ then    
     \be\label{property-1}
     \lim_{(p,\,\tau)\rightarrow P^+}t'= +\infty
      \ee
     where $t'$\  is the time of the next following zero of the solution $x(t;\,p,\,\tau)$. 
     %\hspace{10mm}\lim_{|X^-|\rightarrow +\infty}\tau''= -\infty
     Similarly, 
     \be\label{property-2}
     \lim_{(p,\,\tau)\rightarrow P^-}t''=-\infty
     \ee
     where $t''$\ is the time of the last previous zero of $x(t;\,p\,,\tau)$ for $(p,\,\tau)\in R^-$. 
     The properties (\ref{property-1})\ and (\ref{property-2})\ will play an important role.

%the last previous zero  $\lim_{|X^-|\rightarrow +\infty}\tau'= -\infty$\ (see~\onlinecite{alekseev}, proposition 8). 
%     Later on this property is used and plays an important role.\\ 

     If $e_0=0$\, $D=I$\ and $h^{\pm}$\ coincide with the energy integral ($h^+=h^-=H_0$). The set $R^+=R^-$\ is the circle $|p|\le (2\,V(\infty))^{1/2}$. 

     The return map is derived from the Hamiltonian system (\ref{def:hamiltonian}). The integral $\oint\omega$\ of the Poincar\'e-Cartan invariant $\omega=p\,dx\,-\,H(p,\,x,,t)\,dt$\ assumes the form $\oint_{c} (-p^2/2)\,d\tau$\ on the section 
      $\Gamma$, $c$\ is the contour of a bounded area. It follows that $D$ is area preserving in $\Gamma$ ($p$\ and $\tau$\ are polar coordinate in $\Gamma$).

%     If $e_0=0$\ we have $D=I$\ and the function $h^{\pm}$\ coincide with the energy integral ($h^+(p,\,\tau)=h^-(p\,\tau)=h(p)$); the set $R^+=R^-$\ is the circle $|p|\le (2\,V(\infty))^{1/2}$.\\  
%     \comment{We suppose that when $e_0\neq 0$\ the set $R^+$\ is still contained in a disk.}\\   
     Whenever\ $e_0\neq 0$\ the bounded regions $R^+$\ and $R^-=D(R^+)$\ have the origin in common and have the same area. Therefore the curve $P^+$\ can lie neither wholly inside nor wholly outside $P^-$.
     It follows that\ $P^+\cap P^-$\ is\ non-empty.

%     It is important to note that the character of the solution $x(t;\,p,\,\tau)$ is not affected as $t \pm\infty$. In fact in~\onlinecite{alekseev} (theorem 2) is established that almost all solutions are or 
%     oscillatory, or parabolic for both $t\rightarrow +\infty$\ and $t\rightarrow -\infty$\ or hyperbolic  for both $t\rightarrow +\infty$\ and $t\rightarrow -\infty$; the set of solution that change character have 
%     Lebesgue measure zero.\\  
     We suppose there exists\ $A\in(P^+\cap P^-)$\ at which\ the tangents to $P^+$\ and $P^-$\ are distinct; $A$\ is called 'regular' point. 

\subsubsection{The structure of the scattering function.}
 Following the Ref.~\onlinecite{Eckelt-Zienicke}, we catch the pattern of the scattering map through the map $D$\,. In the section $\Gamma$ we give a construction of the sets of orbits with a determined number of crossings through $x=0$. The structure of the sets shows the properties that we have pointed out in the analysis of the hierarchy in the scattering function.  
%      The idea is adapted from Ref.~\onlinecite{Eckelt-Zienicke}.\\
%     a qualitative picture that catch the worthy items of the involved pattern of the scattering map.\\

      We have already noted that the  hierarchy of the regular intervals is established by the number of crossings of $x=0$. This prompts to think that $D$\ and its iterates should play a role.
     \begin{figure}[!h]
         \begin{narrow}{-0cm}{0cm}
           \centering
            \includegraphics[width=0.45\textwidth]{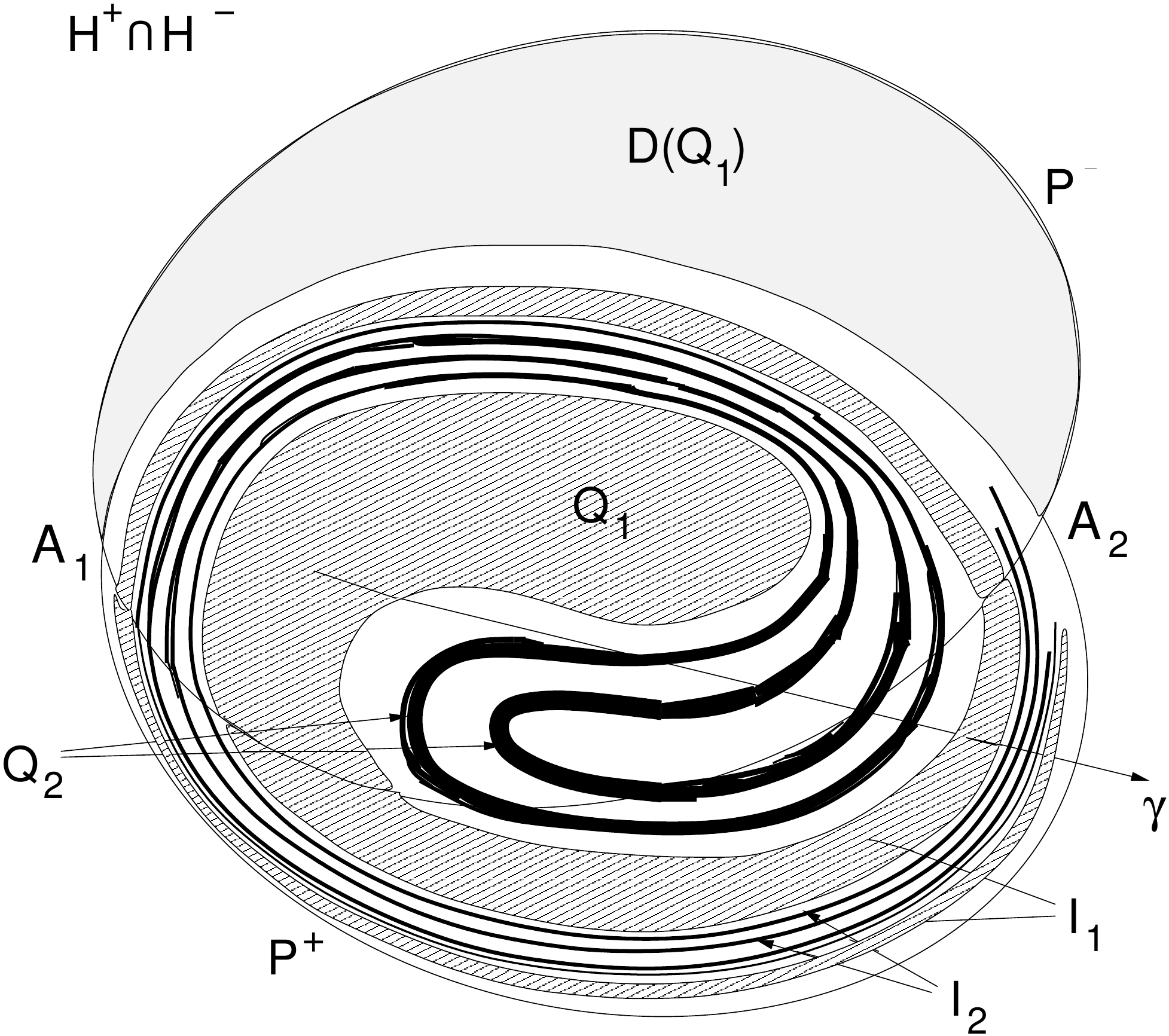}
           \caption{The set $Q_1$\ is the pre-image of $R^-\cap(P^+\cup H^+)$: it is a spiral winding on the boundary $P^+$. $Q_2$\ 
     is the pre-image of $R^- \cap Q_1$\ and it consists of an infinite set of double spirals; two representatives are shown. The sets 
     $I_i=Q_i\cap(P^-\cup H^-)$, $i=1,\,2$, are also displayed. The line $\gamma$\ represents a set of initial conditions 
     ($x_0=0,\,p_0$)\ of orbits that get a map of type of fig. \ref{fig:scattering-function}.
%it is displayed one representative.
            \label{fig:hierarchy}}    
         \end{narrow}
       \end{figure}
      Consider the sets
      \be
        \begin{split}
         Q_n&=\{(p,\,\tau)\ :\ D^{k}(p,\,\tau)\in R^+\ \ \ k=0,\,1,\,\ldots\ n-1\ , \\
              &\quad  \ D^n(p,\,\tau)\in (P^+\cup H^+)\}
        \end{split}
      \ee
       i.e. $x(t;\,p,\,\tau)$\ escapes to a state of scattering after  at least $n+1$\ zeros. $Q_n$\ is non-empty.       
%      i.e. $x(t;\,p,\,\tau)$\ experiences its first zero at $t=2\,\pi/\nu$ and escapes to a state of scattering after it has $n+1$\ zeros. $R^+_n$\ is non-empty.\\ 
       We want to understand the arrangement of the sets $Q_n$\ in the domain\ $R^+$ 
%(see fig.\ref{fig:return-map}).\\    

       We start by noting that\ $D(Q_1)=R^-\cap (P^+\cup H^+)$\, (see fig. \ref{fig:hierarchy}). The arc $A_1A_2$\ of the boundary of\ $R^-\cap (P^+\cup H^+)$\ intersects $P^-$\ transversally at the regular points\ 
       $A_1$\ and $A_2$; therefore its pre-image is a double spiral in $R^+$\ as coming from $P^+$\ as approaching to it. The double spiral separates $Q_1\subset R^+$\ from $R^+\setminus Q_1$.
       It is straightforward to see that $D(Q_2)=Q_1\cap R^-\subset (R^+\cap R^-)$. It follows that $Q_2\subset (R^+\setminus Q_1)$.\ The set $Q_1\cap R^-$\ consists of an infinity of components that 
      accumulate to $P^+$. Except, possibly, for a finite number, each of the components   
       meet $P^-$\ transversally on both ends. As above, each stripe is mapped by $D^{-1}$\ to a loop spiraling twice against $P^+$ (from the property (\ref{property-2})) and that, again, crosses 
       $P^-$\ transversally. Hence $Q_2$\ is an infinity of double spirals (see fig.~\ref{fig:hierarchy}). The tail of $Q_1$\ is located between the branches of each double spiral; the sequence of these double spirals 
       accumulates to the boundary of $Q_1$.
 
       This construction is repeated indefinitely. We find that $Q_n=D^{-1}(Q_{n-1}\cap R^-)$,\ $n\in\mathbb{N}$,\ is an infinite set of double spirals and each one of these is the image, under $D^{-1}$,\ of a strip in 
       $R^-\cap R^+$,\ transversally met 
       by $P^-$. Between the two tails of a double spiral of $Q_{n-1}$\ there is an infinity of double spirals of $Q_n$. 
       In fact, we next show that $Q_n\subset R^+\setminus (\cup_{m=1}^{n-1}Q_m)$. Let us assume $Q_n\cap Q_{n-k}\neq\emptyset$, $k=1,\,2,\,\ldots ,\,n-1$. We take $D^{n-k}(Q_n)$\ and $D^{n-k}(Q_{n-k})$\ and we conclude 
       that\  $(Q_k\cap R^-)\cap ((P^+\cup H^+)\cap R^-) \neq \emptyset$\,. This conclusion is false because $Q_k\subset R^+$.
Finally, $Q_{n-1}$\  and\ $Q_n$\ are dis-joined\ because they are images, under $D^{-1}$,\ of dis-joined set.
 
       We focus on the set $I_n=Q_n\cap (P^-\cup H^-)$,  i.e. we consider solutions with a finite number of zeros. They are parabolic or hyperbolic for $t\rightarrow +\infty$\ and for $t\rightarrow -\infty$.\\ 
%They all are for all time either parabolic or hyperbolic (see~\onlinecite{alekseev}, theorem 2).\\ 
       For all $n$\,, $I_n$\ is a subset of $R^+\setminus R^-$. From the above construction for all $n\in\mathbb{N}$\ between two stripes $I_n$\ there is an infinity of stripes  $I_{n+1}$.

\begin{figure}[!h]
         \begin{narrow}{-0cm}{0cm}
           \centering
            \includegraphics[width=0.42\textwidth]{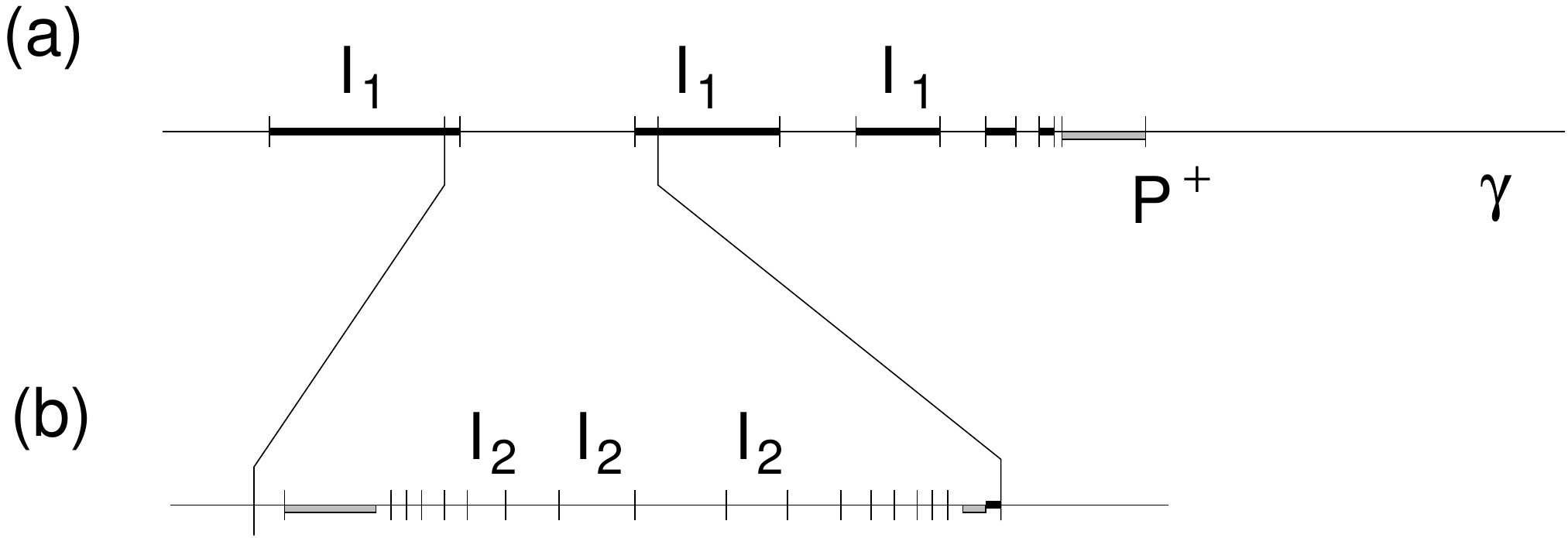}
           \caption{The structure of the intersections with a line $\gamma$, transversal to the boundary $P^+$. In $(a)$ it is shown the set $\{I_1\}$ of intervals cut out on the spiral $Q_1$. The plot $(b)$ displays 
            the intervals $I_2$ cut out on the infinite number of double spirals between two branches of the tail of $Q_1$.\ \label{fig:hierarchy-1}}    
         \end{narrow}
       \end{figure}

       We are now able to explain the structure of the scattering maps (like  fig.~\ref{fig:scattering-function}).
       We carry out the fig.~\ref{fig:scattering-function} by fixing the initial position ($x(0)=0$)\ and scanning an outgoing variable (the energy $H_0$) as a function of the initial velocity. In the plane $\Gamma$ the organization of the scattering map (see\ figures \ref{fig:crossing-zeros}-\ref{fig:N_c--vs--x_0-v_0-1.eps}) is connected to the pattern of the intersections of the set 
$\cup_nI_n$\ with the radial line $\tau=0$ (see fig. \ref{fig:hierarchy-1}). 
%is given when we consider the pattern of the intersections of the set $\cup_n I_n$\  with a radial line ($\tau =0$\ is fixed; see fig. \ref{fig:hierarchy-1}). 
In fig. \ref{fig:hierarchy} the radial line $\tau=0$\ is denoted by  $\gamma$.

       Note that the construction of fig. \ref{fig:hierarchy-1}\ is analogous to the construction of a Cantor set. In the first stage one removes a countable number of intervals $I_1$ that accumulate to $P^+$. In the second stage from each remaining interval a countable number of subintervals converging to both the ends of the interval are removed, and so on.
%%       On account of the above remarks we may state that almost everywhere the scattering map is regular; there is an uncountable set of points 
%with measure zero 
%%       where the orbit is captured and forever evolves close to  $x=0$.\\            
%%%       We conclude that a regular interval and an irregular one are not truly different but the sequence of the sets $I_n$\ occurs over all little scales of the initial condition; each graphic representation of the
%%%       scattering map allows to distinguish until a 'smallest' $I_n$-interval.          
       The return map provides a construction of an arrangement of sets in the phase space that strongly reminds us the hierarchical structure of the scattering function. The analysis is qualitative but effective.

       The scattering map could play a more important role. In the Refs.~\onlinecite{alekseev} and \onlinecite{Eckelt-Zienicke} the authors analyze models with scattering by a one dimensional, periodically oscillating potential well. They construct a set $\Lambda$\ which is invariant under $D$\ and the dynamics of $D$\ in $\Lambda$ is topologically conjugated to a symbolic dynamics; thus they conclude that their models show chaotic scattering. The potential is attractive with respect to the origin for all times;  by means of this property they 
       control the dynamics in the vicinity of the parabolically escaping orbits. This control is essential to establish that the return map is locally a horseshoe map and that $\Lambda$\ is a (non compact) Cantor set.

   \section{The fractal structure of the delay time function and the algebraic decay law.}
     \begin{figure}[!h]
         \begin{narrow}{-0cm}{0cm}
           \centering
            \includegraphics[width=0.45\textwidth]{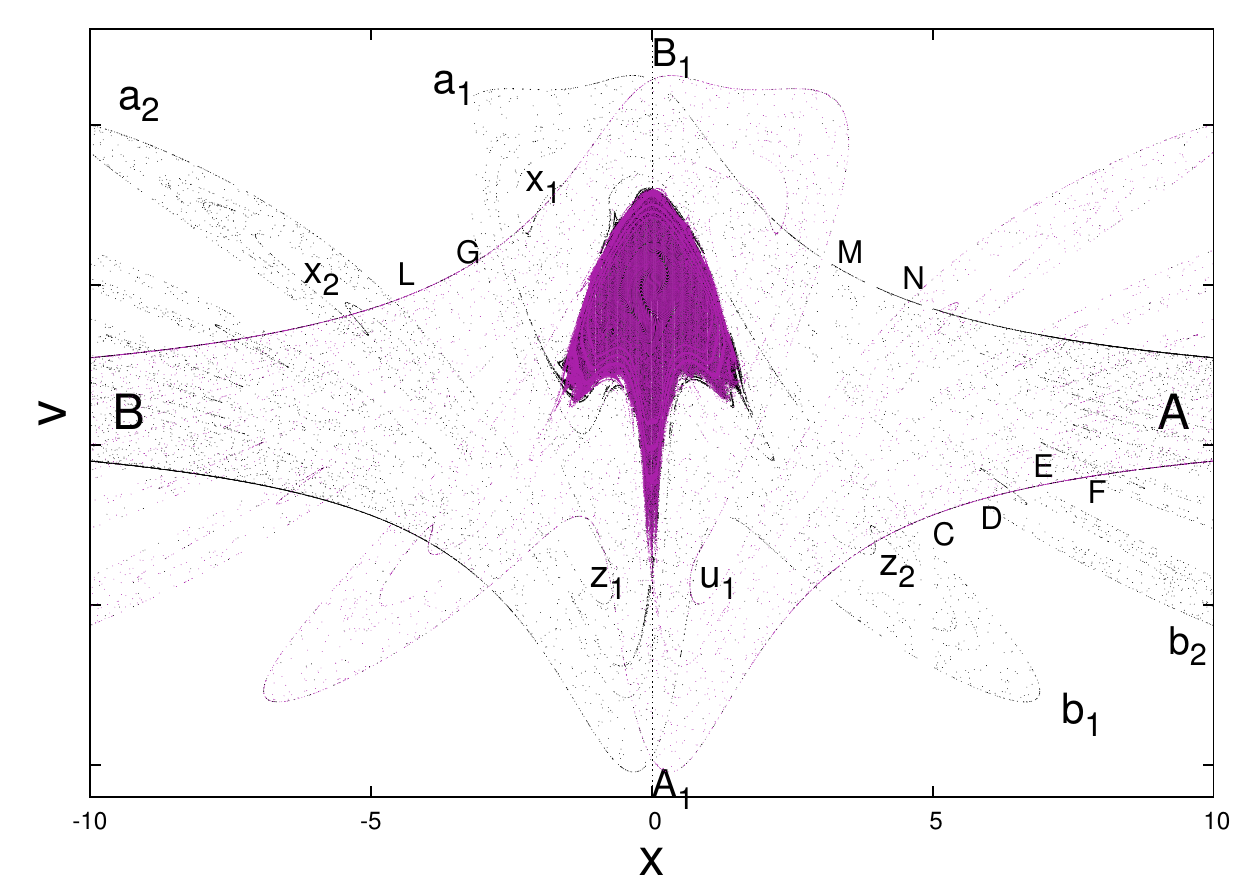}
           \caption{There are shown  the stable manifolds (black) and the unstable ones (red)\ of the outermost fixed points $A$\ and $B$. They are determined through the method of sprinkler.  
            The fundamental region $\mathcal{A}$ is the 'rectangle' $AA_1BB_1$.\ $a_i$\ and $b_i$\, $i=1,\,2$, are the stable tendrils of $A$\ and $B$, respectively. The points $x_i$,\ $z_i$, $i=1,\,2$\ mark the tips of
            the stable gaps of first and second order. The point $u_1$\ is the tip of the first order unstable gap.
            \label{fig:stable-manifold}}    
         \end{narrow}
       \end{figure}
       \begin{figure}[!h]
         \begin{narrow}{0.2cm}{0cm}
           \centering
           \includegraphics[width=0.45\textwidth]{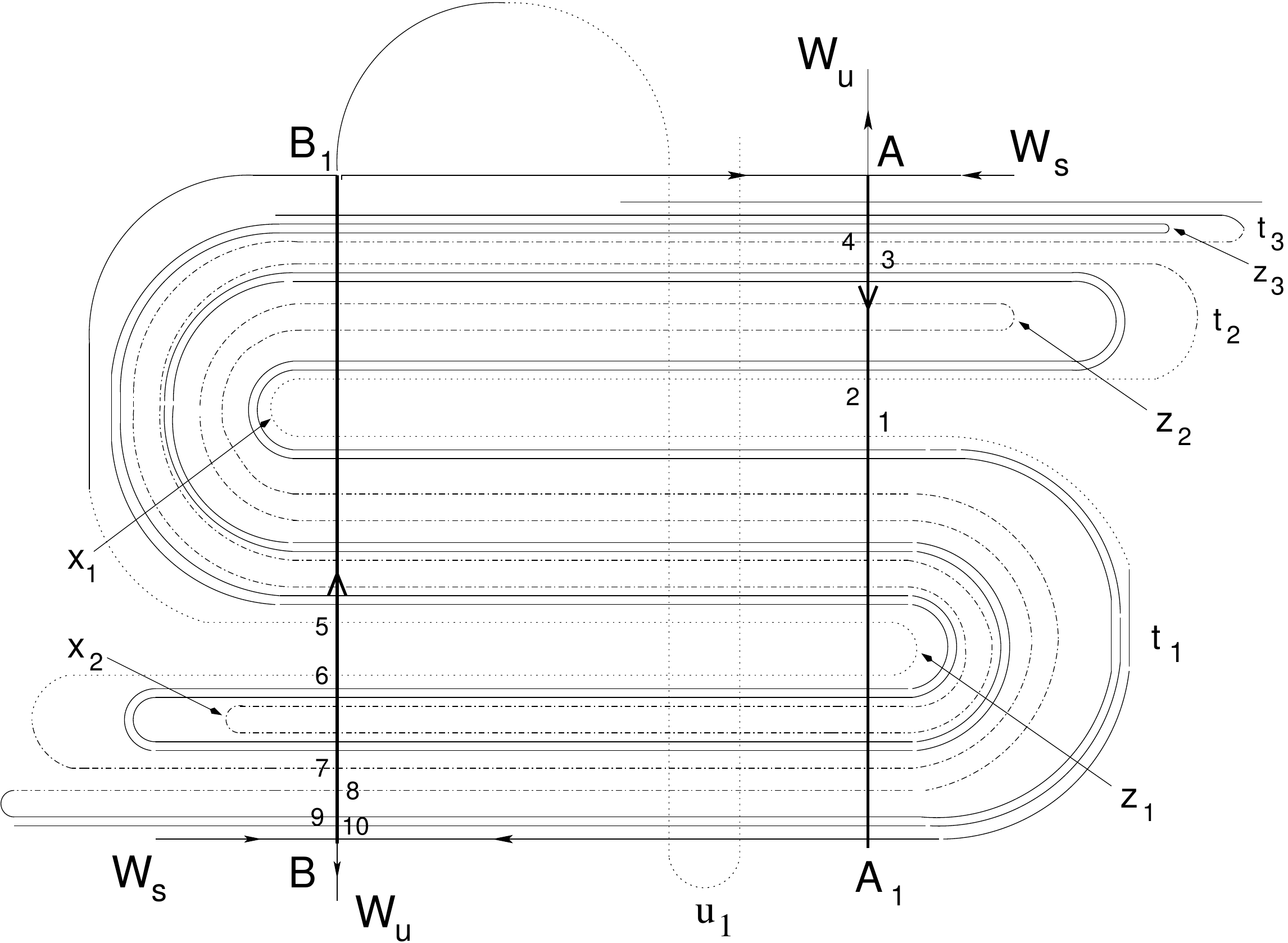}
           \caption{\label{fig:horseshoe} Schematic plot of a complete horseshoe. The outermost fixed points are labeled $A$\ and $B$. The rectangle $AA_1BB_1$\ is the fundamental region $\mathcal{A}$. There are displayed  
            gaps up to order three. The point $x_i$\ is the tip of the stable gap of $B$\ of $i-th$\ order. The point $t_i$\ is the tip of the stable tendril of $A$\ of $i-th$ order. $z_1$\ and $u_1$\ label, 
            respectively, the tip of stable and unstable first gap of $A$.}    
         \end{narrow}
       \end{figure}

\subsection{The fundamental region and the invariant manifolds.}
     We are addressed to give a connection between the delay time of a trajectory and the number of steps that it performs in a given region of the phase space (in this section we deal with the periodic driver ($f=f_2=\sin(\nu\,t)$). In order to do this we use some ideas  
     of Ref.~\onlinecite{Jung-Lipp-Seligman} and Ref.~\onlinecite{Emmanouilidou}.
%     The fundamental region $\mathcal{A}$\  
The considered region is a part of a phase space which contains the invariant set of the scattering process. Its boundaries are traced out by the segments of the invariant manifolds corresponding to the outermost fixed points in position space. Some non-hyperbolic areas may be found in the region.
      
%(in the figure (\ref{?}) is the rectangle $AA_1BB_1$) is that part of the Poincar\'{e} section where the dynamics is hyperbolic but some subregions; the map acts and gives rise 
%     to the invariant set. The boundary of $\mathcal{A}$\ 
%     is traced out by the segments of the invariant manifolds corresponding to the outermost fixed points in position space.\\ 
%     In this section we deal with the periodic driver ($f(t)=f_2=\sin(\nu\,t)$).\\
     We determine the invariant manifolds (the stable manifold $W_s$\ and the unstable one $W_u$)\ by means of the 'sprinkler algorithm'  that was introduced in Ref.~\onlinecite{KG85}. The method consists in considering a distribution of initial conditions from a region that contains the chaotic set. Each trajectory is followed until a given time $t$\ large enough. If the initial point is close enough to the stable manifold it spends a time larger than $t$ to escape from the given region along the unstable manifold. Therefore the initial points, that still stay in the region at $t$,\ form the stable manifold and the set of points at time $t$\ forms the unstable manifold.
 
% The construction is carried out on the Poincar\'{e} section for $\nu\,t_k=\pi/2\,+\,k\,2\,\pi$.  
The fig.~\ref{fig:stable-manifold}\ gives the result of the numerical calculation; we use a grid of $20000\times 6000$\ initial points: the stable manifolds are shown in black and the unstable
     ones are shown in red. We focus on the diamond-shaped area  $AA_1BB_1$ that we name fundamental region and denote by $\mathcal{A}$.     
     The points $A$ and $B$ are the outermost saddle points respectively at $x >0$\ and $x < 0$. 
     There is an elliptic fixed point close to the origin; it is encircled by KAM tori, secondary elliptic points and other possible invariant structures that are remnant of the quasi-periodic case. Altogether 
     these structures determine a region of phase space where the dynamics is not hyperbolic.

     The Hamiltonian (\ref{def:hamiltonian}) is not invariant under the transformation $x\rightarrow x'=-x$; therefore the underlying invariant structure seen by the particle is different when it comes from opposite 
     directions.        
%and so the resultant pattern does not hold the left-right symmetry. 
     However the Hamiltonian is invariant under the composition of $t\rightarrow -t$\ and $x\rightarrow -x$; it follows that, in the resultant pattern of 
     fig.~\ref{fig:stable-manifold}, we convert each stable manifold to an unstable one and vice versa by letting $x\rightarrow -x$.  
%if we consider both the space inversion and the time reversal transformation, which means that we convert each stable manifold to an unstable one by letting $p\rightarrow p'=-p$\ and $f\rightarrow -f$,\ 
%     then the Hamiltonian field transforms like: 
%     \be\nonumber
%       \left (\begin{array}{c}   
%                \partial_pH(p,\,x,\,t)\\
%               -\partial_xH(p,\,x,\,t)  
%              \end{array}  
%        \right )        
%        \longrightarrow   
%        \left (\begin{array}{c}   
%               - \partial_pH(p',\,x',\,t)\\
%               -\partial_xH(p',\,x',\,t)  
%              \end{array}  
%        \right )
%     \ee    
%    It follows that the resultant pattern (see fig.~\ref{fig:stable-manifold}) is symmetric under the composition of $t\rightarrow -t$\ and $x\rightarrow -x$. This
%    property is satisfied by the pattern of fig.~\ref{fig:horseshoe}.
   
    The intertwined structure of fig.~\ref{fig:stable-manifold} is obtained by starting from the segments of the manifolds $W_s$\ and $W_u$\ that make up the boundaries of $\mathcal{A}$.
    The homoclinic/heteroclinic intersection points between $W_s$ and $W_u$\ belong to the chaotic set and the topology of the intersections 
    determines the structure of the outermost component of the set.

    The fig.~\ref{fig:horseshoe}\ gives a schematic plot of the intersections set of the invariant manifolds of $A$ and $B$  (we do not consider chaotic structures owing to inner fixed points). It is the known 
    construction of the Smale horseshoe: the fundamental region $\mathcal{A}$ (whose boundaries $AA_1BB_1$ are given by segments of the invariant manifolds of the outermost fixed points $A$ and $B$) is stretched and 
    folded back onto itself. We fold $\mathcal{A}$ two times because we have two fixed points.   

    We now describe in more detail the construction of fig.~\ref{fig:horseshoe}.  
%    Even more holds: 
%    the topology of the intersections is determined by the intersection pattern between all the tendrils of the stable manifolds and the 'local' segments of the unstable manifolds, i.e. the branches $AA_1$\ and $BB_1$ in 
%    the figure (\ref{fig:stable-manifold}). This fact can be easily understood if we think  that each intersection (homoclinic, heteroclinic) point in $\mathcal{A}$ is the image-pre-image of an intersection point 
%    of $W_s$\ with the local segments.  
    The segments of the boundaries of the fundamental region (the unstable segments $AA_1$\ and $BB_1$ and the stable ones $AB_1$ and $BA_1$) are called tendrils of zero order. 
%    Using the equations of motion of the Hamiltonian (\ref{def:hamiltonian}) 
We propagate the segments of the stable manifolds backward in time and the segments of the unstable manifolds forward. We focus on the stable 
    manifold of the fixed point $B$\ and the local segment of the unstable manifold of $A$. 
    The pre-images (images) of the stable (unstable) tendrils of zero order are tendrils of first order (the segment $A_1t_11$\ of $W_s$\ is the tendril of order one);     generally the tendril of order $n$ is mapped to that of order $n+1$ (the arc\ $2t_23$\ of $W_s$\ is the tendril of order two and it is the pre-image of $A_1t_11$).
      
    In between these segments of $W_s$\ forming the tendrils there are other segments $g_n^B$\ of $W_s$\ going across the fundamental region; they are called gaps of order $n$. $g_1^B$ is the arc $1x_12$.\ The points of $g_1^B$\ are mapped to the segment $3x_24$, i.e. the stable second order gap $g_2^B$. Generally the pre-image of $g_n^B$\ is the gap $g_{n+1}^B$.
    We may similarly proceed with the invariant manifolds of $B$; the segment $5z_16$\ is the first gap $g_1^{A}$\ of $W_s$\ of the fixed point $A$; $7z_28$\ and $9z_310$\ are the gaps $g_2^A$\ and $g_3^A$, respectively, and so on.

   In this construction we use the invariant manifolds of the fixed points on the corners of $\mathcal{A}$ ($A$\ and $B$). This choice implies a favorable property which does not hold in the case of the invariant manifolds of periodic points lying in the interior of $\mathcal{A}$. In fact the gaps which are cut into $\mathcal{A}$ by the manifolds of $A$ and $B$ are regions of $\mathcal{A}$\ which do not cover the invariant set;
    no higher level segment of $W_s$\ will ever enter such gaps.
     Furthermore it holds: 
    the topology of the intersections is determined by the intersection pattern between all the tendrils of the stable manifolds and the 'local' segments of the unstable manifolds, i.e. the branches $AA_1$\ and $BB_1$ in 
    the fig.~\ref{fig:stable-manifold}. This fact can be easily understood if we think  that each intersection (homoclinic, heteroclinic) point in $\mathcal{A}$ is the image-pre-image of an intersection point 
    of $W_s$\ with the local segments. 
%    The gaps cut strips which are free of further segments of $W_s$  out of $\mathcal{A}$. We can conclude that if the invariant chaotic set is given it is contained in the complementary  set in $\mathcal{A}$\ 
%    of the gaps.\\
   
%    The intersections of the first order tendril of the stable manifold of a fixed point ($A$, $B$) with the boundary of the fundamental region (zero order segments of the invariant manifolds)  yield the first order gaps 
%    $g_1$ (see fig. ?); generally, a gap of order $n$, $g_n$,\ is one of the strips of the stable tendril of order $n$ that go across $\mathcal{A}$. A gap cuts a strip, which is empty of further segments of $W_s$, out of 
%    the the fundamental region.\\

\begin{figure}[!h]
      \begin{narrow}{0.cm}{0cm}
           \centering
           \includegraphics[width=0.42\textwidth]{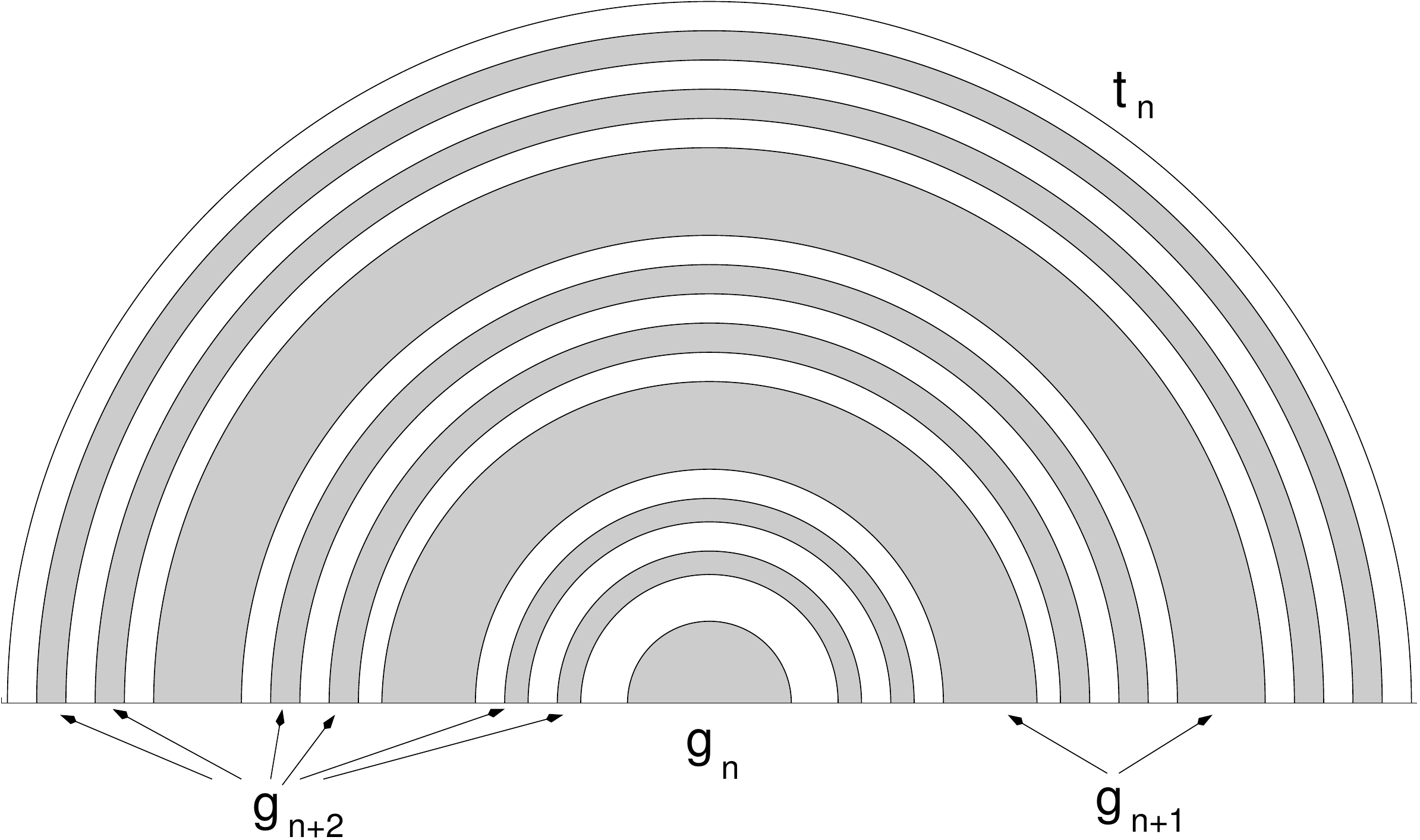}
           \caption{The plot displays the first few stages of the development of the pattern of gaps in a tendril of order $n$.\label{fig:tendril}}    
       \end{narrow}
     \end{figure}

    The horseshoe construction is complete when the tip $u_1$\ of the first order gap of the unstable manifold reaches the other side of the fundamental area (the schematic plot of fig.~\ref{fig:horseshoe} shows a 
    complete horseshoe). In this case a stable gap of order $n$\ reaches into all tendrils up to the order $n$. 
    The fig.~\ref{fig:tendril} shows a partial development of the pattern of gaps in a tendrils of order $n$. The innermost part is a gap of order $n$. Between this middle gap and the boundary of the tendril there are two 
    gaps of order $n+1$. On the next level there are two gaps of order $n+2$\ in between each pair of adjacent gaps of lower order or between the gaps of lower order and the boundary. This scheme continues to higher level. 
    In a tendril of order $n$ we find $2 \times 3^{n_1-1}$\ gaps of order $n+n_1$.  
    Between any two adjacent gaps of the same order we find gaps of order arbitrarily high; hence in each neighborhood of a gap 
%of order $n_1 > n$\ 
we find gaps of all orders high enough. In the area enclosed between a tendril and the local segment the gaps fill a subset of null 
    measure. The resultant structure is a set of segments of $W_s$\ along the 
    direction of the tendril whose sections in the transverse direction are Cantor sets. This structure is closely connected to the invariant set in the fundamental region.
%The pattern of the intersection points along the zero order tendrils is a Cantor set.\\   
%    Relating to a scattering problem the worth of the gaps consists in the following mechanism.\\ 

    The link between the gaps and the scattering problem is due to the following mechanism.
    The tendrils reach out into the incoming asymptotic region. We consider a particle that comes from far away and approaches the local segment and lies in a gap of a tendril of order $n$. Except for a fractal 
    set of lines (resembling the invariant set) the particle is mapped to a point of a gap in $\mathcal{A}$. In fact, by the construction of the horseshoe, any point in the stable tendril of order $n$\  is mapped 
    in the stable tendril of order $n-1$.  The first order tendril of the stable manifold of $A$\ ($t_1^s$) is mapped in the first gap of the unstable manifold of $B$ ($u_1$); more specifically, a point in 
    a stable gap of order $n$\ ($g_n^s$) in $t_1^s$\ is brought into $u_1\cap g_{n-1}^s$\,.  Points of a gap are brought out of the fundamental region after a finite number of iterations 
    of the dynamical map: the stable gap $g_n^s$\ is mapped to $g_{n-1}^s$\ and the image of $g_1^s$\ is the first order unstable tendril of $B$. At this point the particles leaves out of $\mathcal{A}$; in fact  an unstable 
    tendril of order $n$\ is mapped to an unstable tendril of order $n+1$, and so on. Finally, except for a set of zero measure, a scattering trajectory experiences a finite number of steps  inside the fundamental area
    whenever it approaches the local segment of the unstable manifold along a gap in a stable tendril.

     From these considerations we assert that the arrangement of singularities of a scattering function should be related to the structure of the chaotic invariant set and then to the sequence of tendrils along the local segment 
    of the unstable manifold. In this sense we can reconstruct the structure of the horseshoe from the measure of the scattering functions, i.e. from asymptotic measurements.   

%It follows that the gaps correspond to areas of the fundamental region which no point of the invariant chaotic set belongs to.         
%zero order segment of the unstable manifold of the fixed point $B$\ yields the first order gap $G_1$ (see fig. ?)     

%     In our case the outer fixed points are located at $x=+\infty$\ and $x=-\infty$. They are parabolic and each of them yields a stable manifold and an unstable one. These manifolds give the boundaries of the fundamental 
%     region. The Poincar\'{e}\ section is determined for $\nu\,t=\pi/2$.      

%---------------------------------------------------------------------------------------------

     \subsection{The scattering function and the structure of the invariant sets of the outermost fixed points.}
\begin{figure}[h]
        \begin{center}
          \includegraphics[width=0.45\textwidth]{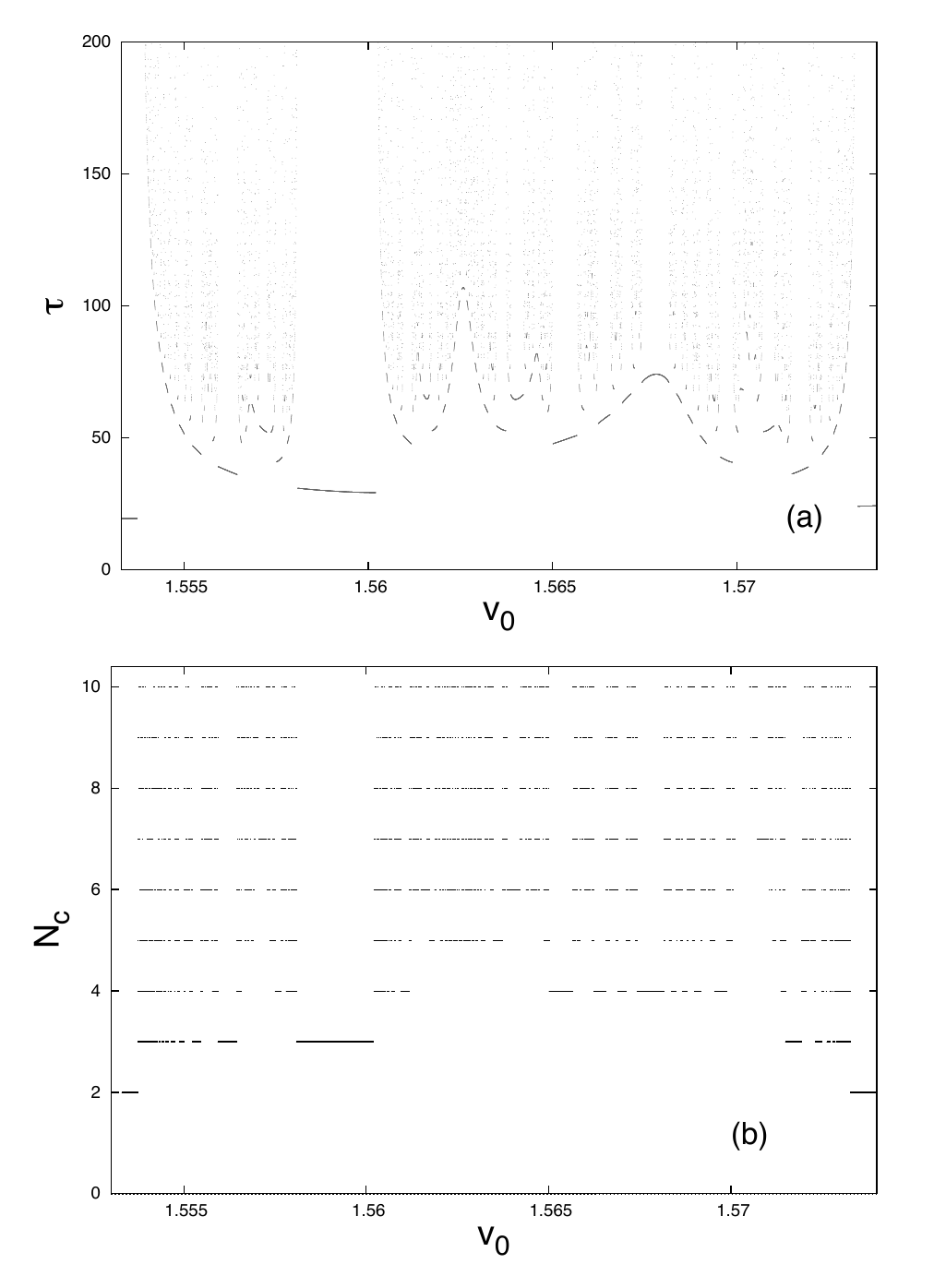} 
          \caption{The plot $(a)$ displays the delay time scattering function for the same line of initial conditions of $(b)$\ fig. \ref{fig:scattering-function}. The plot $(b)$  is the number of crossings of $x=0$\ in 
           the same conditions of $(a)$. \label{fig:time-delay}}
        \end{center}
      \end{figure}

     We know that the set of singularities of a scattering function shows how the line of the incoming particle asymptotes pierces the bundle of stable manifolds of the chaotic invariant set. 
     However we may state that the main properties of a scattering function are dominated by the component of the invariant set due to the outermost fixed points.\\   
     We focus on the delay time scattering function; we keep the initial position fixed and vary the initial velocity and for the corresponding trajectories we monitor the time interval $\tau$\ between the first and the 
     last zero. 
     The plot $(a)$\ of fig. \ref{fig:time-delay} shows a typical pattern. The structure of the delay time closely corresponds to the structure of the $N_c$ scattering function (compare  $(a)$ with $(b)$): 
     \begin{itemize}
     \item [-] an interval with a given number of zeros is related to a continuous branch of the delay time function;
     \item [-] each regular interval of $\tau$\ is accumulated by a sequence of regular intervals for $\tau$\ going to arbitrary large values. 
     \end{itemize}
     We are mainly interested in the singularity structure of the scattering functions. For this particular purpose the information provided by $\tau$\ is the same as the time interval which the particle spends 
     in the fundamental region. 
     We have focused our attention on the features of the dynamics of particles approaching the fundamental region through the local unstable segment. Our considerations drive us to 
     affirm that the singularities structure in $\tau$ is correlated with the structure of intersections of $W_s$\ of a fixed point with the local branch of $W_u$ of the other fixed point, i.e. the pattern of fig. 
     \ref{fig:tendril}.                
     
     On account of the structure of gaps in a tendril, we see that, except for a set of zero measure, the trajectories, coming from a line of initial conditions and intersecting the tendril, are mapped far away after they  undergo a finite number of iterates inside the fundamental region. A trajectory that approaches the fundamental region along a gap 
%of order $n$\ then it spends only $n-1$\ steps  inside $\mathcal{A}$. This trajectory  
belongs to a regular interval of the delay time function.
     On the other hand the set of singularities of the delay time function corresponds to the set of points in a tendril that do not belong to any gaps; the trajectory is confined forever inside the fundamental region.

      The graph of $\tau$ possesses the following hierarchical organization. 
%Consider an horizontal line $\tau=\tau_0$\ that moves from $0$ toward increasing values of $\tau$; it cuts out continuous branches of the graph 
%      with $\tau < \tau_0$ so that the structure can be separated into substructures that are similar to the complete one. Again, these substructures have lowest continuous branches; the following cutting eliminates these
%      branches and creates new substructures. We may repeat this procedure many times, i.e. we determine a hierarchical organization.\\
      Consider the connection between $\tau$\ and $N_c$. In the graph $(b)$\ of fig.~\ref{fig:time-delay}\ we remove all continuous branches corresponding to the intervals with the lowest $N_c$; the $\tau$ structure is 
      separated into substructures that are similar to the parent one; each removed subinterval is surrounded by smaller subintervals with a higher value of $\tau$ ('castle-like' structure; see Ref.~\onlinecite{Ruckerl-Jung}). 
      We may iterate the procedure and the number of zeros fixes the level of cutting out; at each step we get a generation of substructures similar to that of the previous generation. 
      In this way we determine a hierarchical organization.  
      It is important to observe that the pattern of gaps in a tendril (fig. \ref{fig:tendril}) displays the same organization. In a tendril of order $n$, whenever we cut a gap of order $n+1$\ away, the pattern is separated 
      into substructures, with gaps of higher order, similar to each other. We may iterate the method for gaps of higher order; the hierarchy follows the order of the gap.        
%      We have direct access to the interaction region. 
      In this way we establish a connection between the pattern of the delay time function and the structure 
      of the $N_c$ scattering function (compare $(a)$\ and $(b)$ of\ fig. \ref{fig:crossing-zeros}).  
%      In the Refs.~\onlinecite{alekseev} and~\onlinecite{Eckelt-Zienicke} it is established that the dynamics of the system can be derived from the dynamics of the return map $D$\ on condition that we restrict in some neighborhoods $U_i$, small enough, of the regular points; there exists an infinite invariant set of saddle points of $D$ in $U_i$, i.e. an infinite set of unstable periodic solutions of $(\ref{def:hamiltonian})$. A deficiency of this way of reasoning is the restriction to small neighborhoods.
      Likewise the qualitative approach that refers the return map, in this section we have examined some features of the scattering function using global considerations on the dynamics (the invariant manifolds of the fixed points $A$\ and $B$). In addition we have not needed to 
      take into account all the chaotic invariant set.

     \subsection{The algebraic decay.}

       \begin{figure}[!ht]
         \begin{narrow}{-1cm}{0cm}
           \centering
           \includegraphics[width=0.45\textwidth]{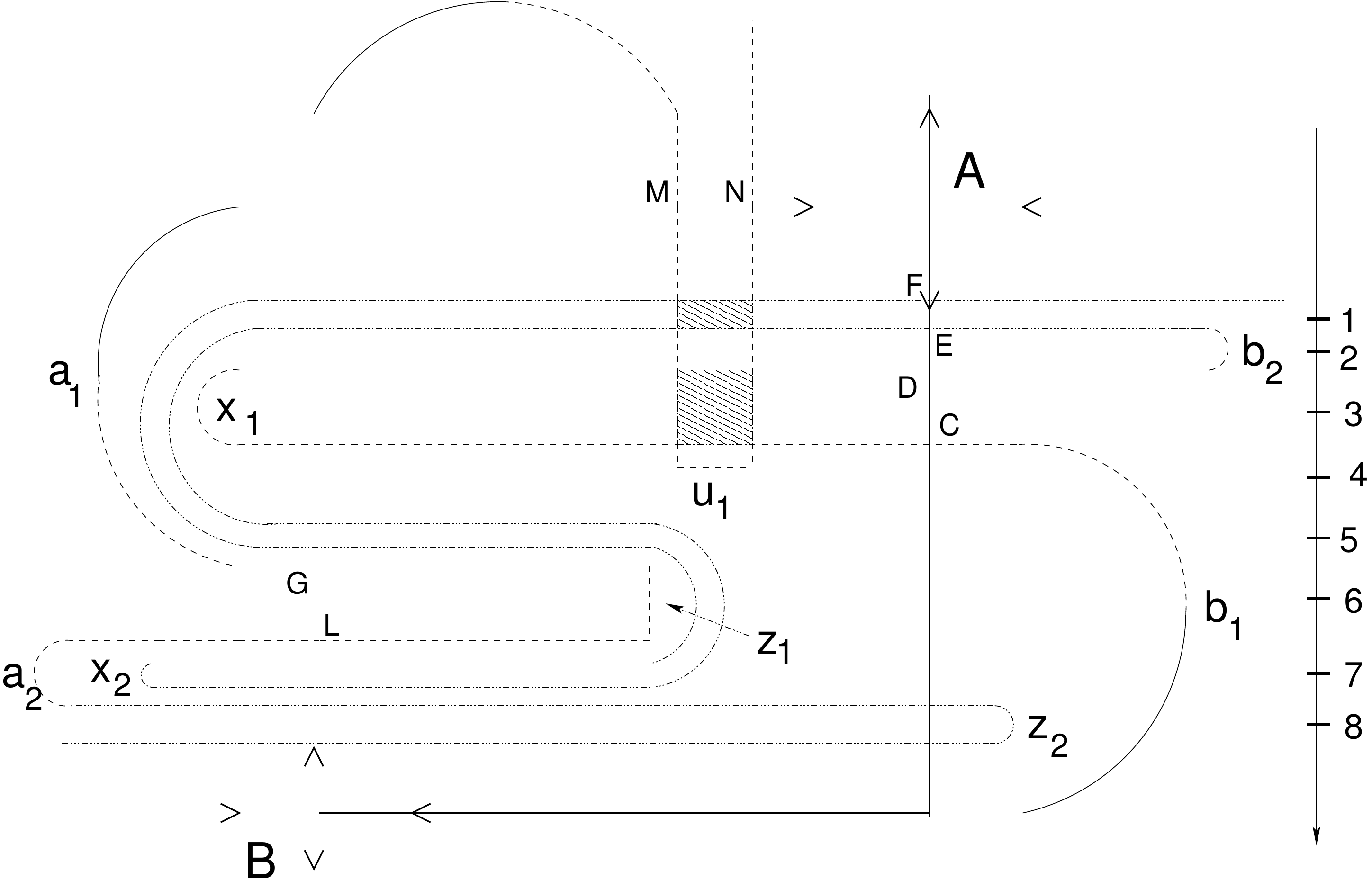}
           \caption{Schematic plot of the scenario in fig. \ref{fig:stable-manifold}. The horseshoe is incomplete and asymmetric; the first order unstable gap (tip $u_1$) does not reach the opposite side of the 
            fundamental area. The numbers in the right column count all gaps up to order two. The horseshoe is characterized by the development parameters $\gamma_A=1/3$\ and $\gamma_B=1$. \label{fig:horseshoe-uncomplete}}    
         \end{narrow}
       \end{figure}

       In section 2,  we have checked an algebraic law both for $N(t)$, the decay with time of the number of temporarily bound orbits,  and for\ $N(n)$, the distribution of zeros. 
       Hyperbolic systems usually exhibit $N(t)$\ to have an exponential decay law \cite{Grebogi-Ott-Yorke,Kovacs-Tell}. The algebraic decay would press to consider our system 
       to be non-hyperbolic \cite{Lai-Ding-Grebogi-Blumel, Meiss-Ott} or, if it were hyperbolic, there should be a special mechanism that makes the decay slower than what is expected when the KAM tori structure is 
       completely broken \cite{Lee,Vivaldi-Casati-Guarnieri}.
       Our problem should fall in with the non-hyperbolic situation. In fact the Poincar\'{e} map shows orbits that curl up tori, i.e. there is a mixture of motion  close to the deformed 
       tori and motion far away from them.

       The presence of KAM surfaces affects the particle motion in the nearby chaotic regions; even if the particle starts in a chaotic region, sooner or later the particle approximates the surface and wanders close to 
       it for a long time; it is the "stickiness effect". The motion in a KAM surface is quasi-periodic and hence, nearby the tori, the Lyapunov exponents in the directions along the surface are zero. Moreover, in our case, 
       the Poincar\'e map is symplectic (the driver is periodic) and, because it is two-dimensional, the Lyapunov exponent perpendicular to the surface also has to be zero. The stickiness effect reduces the transport of orbits initialized in a chaotic region and explains the algebraic decay law instead of the exponential one.
%      In section 2, we have observed that there is a crossover time $\overline{t}$\ such that $N(t)$\ displays an algebraic decay both for $t\le\overline{t}$\ and 
%       $t\ge\overline{t}$\ but the exponent in the first case is less than that one in the other case. We give some considerations through which we gain understanding on this
%       fact. 
%       We know that a Cantori $\mathcal{B}$\ is a temporarily barrier that takes the place of broken KAM curve; a phase point move for a time along $\mathcal{B}$\ before 
%       crossing it. The phase space is divided into the region enclosed by $\mathcal{B}$\ and the region that lies outside $\mathcal{B}$. With regard to our problem the set of initial conditions that generate $N(t)$\ is partly inside $\mathcal{B}$
        In the Ref.~\onlinecite{Lai-Ding-Grebogi-Blumel} it is established that the size of the decay depends on the complexity of the pattern of island chains: the more dominant the island chains in the chaotic region, the 
        slower the escape process.
       
        We have remarked that in the Ref.~\onlinecite{Beeker-Eckelt} an algebraic decay for the delay time function is obtained. Nevertheless in this case the effects of the KAM components are negligible.
%        Beeker et al. claim that the KAM component effects are negligible for the parameter values they choose. Beeker et al. explain the algebraic scaling of the scattering function in the 
%        vicinity of parabolically escaping orbits by a different mechanism than the stickiness of KAM tori. They consider the survival time of the temporarily bound orbits close to the parabolic orbit.
        When we consider $V=V_2$, $V_2 \sim |x|^{-2}$\ for $|x|\gg 1$, the explanation of Beeker et al., applied to our case, supplies a decay law slower than the law we observe. The reasoning of Beeker et al. 
        attaches the most part of the importance to the orbits that spend the most part of time far away from the interaction region; they are trajectories which almost escape after e.g. the $n$-th zero, but are eventually
        recaptured, so that there is a long time elapsing between the $n$-th and the $(n+1)$-th zero.  
%This results to consider the tail for large $t$ in $(a)$\ of\ fig \ref{fig:decay_function--e0_1-x0_0-omega_08-1_55}.\\
        Because of the KAM surfaces, in our case we have to take into account the orbits that run inside the interactions region. Their 'life-time' is shorter than that of the temporarily bound orbits close to the 
        parabolic one; in fact
        the energy absorptions mainly take place around the origin. On the basis of these considerations we expect the algebraic decay to be faster than the scaling determined by the class of orbits in the 
        approximation of Ref.~\onlinecite{Beeker-Eckelt}.  
%%%smaller survival probability of the temporarily 
%%%        bound orbits); there  
%%%        the decay with the time is really slower than the other part.\\  
%%%       survival probability for the temporarily bound orbits, 
%%%        in the vicinity of the parabolically escaping orbits, decays with an algebraic law of time.              
%%%         Beeker et al. obtain an exponential law instead of an algebraic one for the distribution of the zeros $N(n)$. In order to support this result, 
%in Ref.~\onlinecite{Beeker-Eckelt}\ it is 
%they proposed an argument that is essentially based on an effective mixing property of the dynamics. We believe that the mixing property can be brought up only if the KAM scenario is destroyed enough.    
%       The development parameter $\gamma$\ is helpful to give a quantitative characterization of the development stage of the horseshoe construction. It was introduced in Ref.~\onlinecite{Ruckerl-Jung}. 
       In order to give a quantitative characterization of the development stage of the horseshoe construction a parameter $\gamma$ ('development parameter') was introduced in Ref.~\onlinecite{Ruckerl-Jung}. 
       For the system under consideration there are three most effective fixed point: the outermost ones and the inner elliptic fixed points. The development parameter for the corresponding horseshoe is $\gamma = r_n\ 3^{-n}$; $n$\ is the highest order of the gap we consider and $r_n$\ is the order of gap, which the tip of the first unstable gap reaches, when we count the
       gaps from the fixed point (see Ref.~\onlinecite{Jung-Lipp-Seligman}). Note, the value of $\gamma$ does not depend on the choice of the order\ $n$\ we consider.

       The parameter allows to evaluate how deep the  first order gap penetrates into the fundamental area of 
       the horseshoe compared to the complete case; in the development of a horseshoe the parameter $\gamma$\ varies from 0 to 1. 
       A horseshoe is complete whenever the unstable gap of order one reaches the opposite side of the fundamental area (see fig.~\ref{fig:horseshoe}); in this case $\gamma=1$.
%       For the system under consideration the number of fixed points that are the most effective is three: the outermost ones and the inner elliptic fixed point. The development parameter for the corresponding horseshoe
%       is $\gamma = r_n\ 3^{-n}$; $n$\ is the highest order of the gap we consider and $r_n$\ is the order of gap, which the tip of the first unstable gap reaches up to, when we count the
%       gaps from the fixed point (see~\onlinecite{Jung-Lipp-Seligman}). Note, the value of $\gamma$ does not depend on the choice of order\ $n$\ we consider.\\
%       In a complete case the parameter is $\gamma =1$.\\ 

        The parameter is used to qualitatively characterize the topology of the homoclinic/heteroclinic tangle of the outer fixed points. It remains the same when under variations of physical 
        parameters the changes of the topology concern tangencies of  branches of manifolds of level high enough. It is important to point out the relation of $\gamma$ to the topological
        entropy $K_0$.  In Ref.~\onlinecite{Ruckerl-Jung} it is obtained, by numerical computation, that $K_0$\ is a monotonic function of $\gamma$.   
        The parameter ignores the unique aspects of the KAM surfaces; it does not take into account the winding numbers of the elliptic points and  neglects effects coming from the invariant manifolds that penetrate into 
        the surroundings of the KAM islands. However the parameter considers the part of the invariant set which is the most important for the scattering function.

      We focus on the fig.~\ref{fig:stable-manifold}. We observe that the inner non-hyperbolic island obstructs the penetration of the unstable gap in the fundamental area and keeps from the complete growth of the saddle 
      pattern.
      The fig.~\ref{fig:horseshoe-uncomplete} is a schematic plot of fig.~\ref{fig:stable-manifold}; the stable manifolds of $A$\ and $B$\ are plotted up to order two and $W_u$ of the point $B$ is plotted up to order one. 
       The numbers in the 
       right column count all gaps up to order two, consecutively from the top to the bottom.  
       We note that the horseshoe is non-symmetric and hence it is incomplete. Actually the first unstable gap of $A$\ reaches the other side of the fundamental region (it is the mirror image of $x_1$) and 
       the unstable gap $u_1$ does not. The horseshoe is thus described by two development parameters. The parameter value of $W_u$ of $B$  is $\gamma_B=1$. On the other hand, when we consider gaps up to second order 
       ($n=2$) the $u_1$\ ends in the third gap from the top ($r_2=3$), thus leading to $\gamma_A=1/3$.        
       
%       The development parameter gives a partial classification of a horseshoe (in particular, of the pattern of fig. (\ref{?}) \comment{quella ottenuta numericamente}). 
%       The parameter gets universal aspects: in the 
%       development of the horseshoe the hyperbolic component is determined, even though the parameter neglects effects coming from homoclinic tangle of inner fixed points. However it considers the part of the invariant set
%       which is the most important for the scattering behavior. We know that the singularities pattern of the scattering function (in particular the time delay function) allows to approximately reconstruct the 
%       chaotic saddle; the development parameter is a compact way to characterize this approximate description.

       Concerning the scattering problem, the effects of KAM islands and their secondary structures should have little influence on the approximation of 
       the horseshoe reconstruction: they are connected to the penetrations of the invariant manifolds of the outermost fixed points in the secondary structures around the KAM islands.        
       In non-hyperbolic case the KAM component obstructs a full development of the horseshoe.  In contrast to the complete horseshoe (fig.~\ref{fig:tendril})\ in the incomplete case some inner gaps in a tendril can
       disappear (fig.~\ref{fig:horseshoe-uncomplete}). However the non transversal intersections between stable and unstable manifolds due to KAM tori have small effects on the scattering functions 
       \cite{Ruckerl-Jung}, i.e. on the structure of the intersections between the stable manifold and the local segment of the unstable manifold. 
       For $\nu = 0.8$\ and $e_0=1$\  there are tangencies for gaps of order $n\ge 3$. The effects of the tangencies are visible on the scattering functions at 
       the order $2\,n\ge 6$; in fact whenever the $n$-th order stable gap is tangent to the $n$-th unstable gap then the $(n+1)$-th order stable gap intersects the $(n-1)$-th unstable gap and, finally, the $(2\,n)$-th
       stable gap intersects the local segment at the $(2\,n)$-th tendril. 

      \section{Scattering features as the driver amplitude is varied.}
      In this section we fix the initial conditions and vary the parameter $e_0$. 
      We know (see Ref.~\onlinecite{Ding-Grebogi-Hott-Yorke}) that  a scattering system can experience transitions from regular to chaotic scattering as a parameter of the system is varied. The onset of chaotic scattering can be 
      achieved through combination of bifurcation mechanisms: new periodic orbits are added and their stable and unstable manifolds intersect each other. These events cause changes in the chaotic set; as the parameter is varied the hyperbolic component increases and the formation of structures topologically equivalent to a horseshoe may occur.

      We do not address the question of how the chaotic scattering arises as $e_0$\ changes. We only discuss the qualitative behavior of the outgoing energy of a trajectory with given initial conditions. We argue that 
      the main properties are connected to the invariant pattern caused by the outermost fixed points.
      We consider both the periodic case ($f=f_2$) and the finite duration driver ($f=f_1$). 
      The initial conditions are set $x(0)=0$ and\ $\mm{v}(0)=1.049$ ($H_0=0.55$\ at $e_0=0$).
    
\begin{figure}[ht!]
        \begin{center}
         \includegraphics[width=0.45\textwidth]{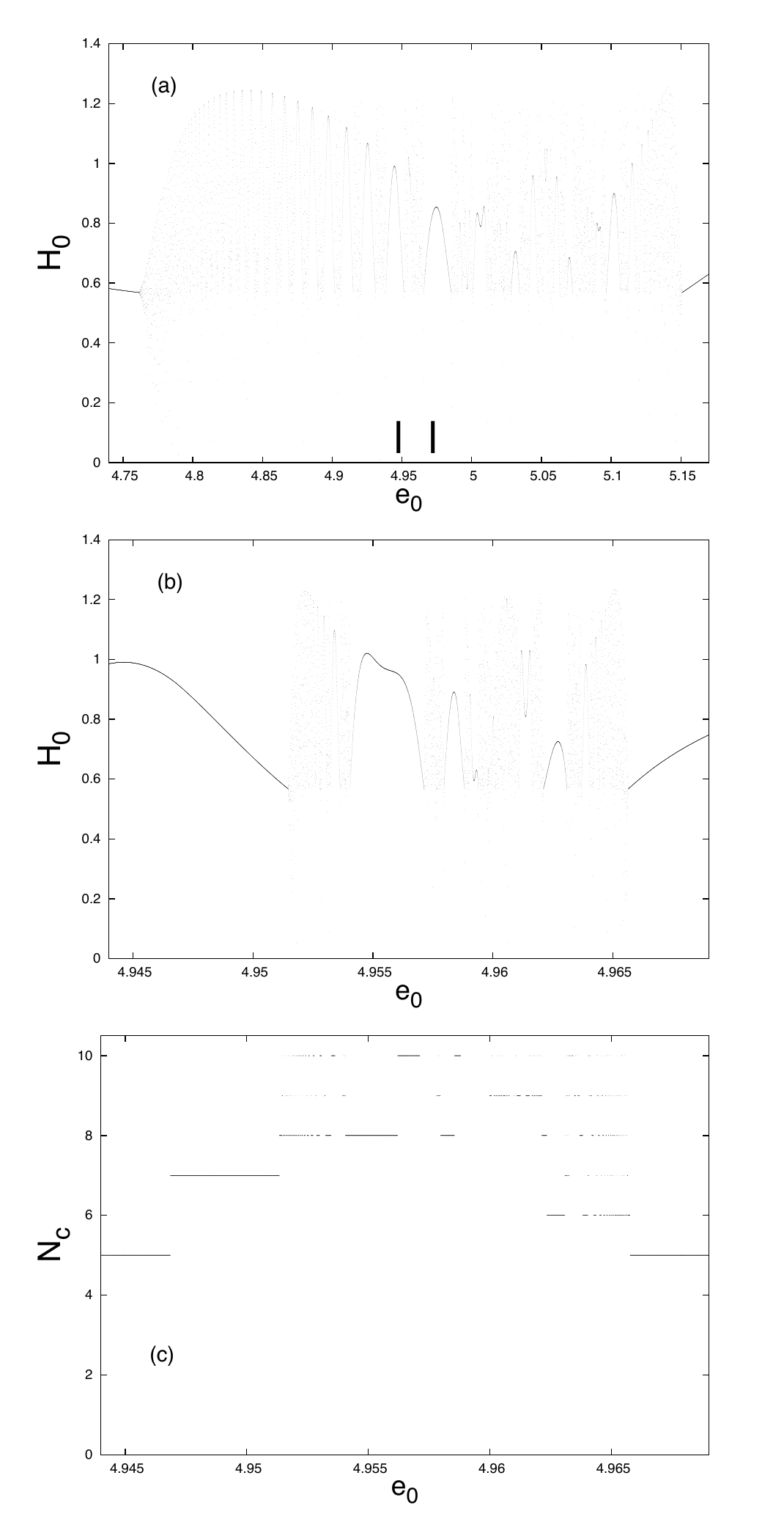}
         \caption{The plot $(a)$\ displays the  asymptotic-out energy ($H_0$)  vs the driver amplitude ($e_0$) for $f=f_1$.  The plot $(b)$ is a magnification of the marked interval of $(a)$.  The plot $(d)$ shows 
           $N_c$ corresponding to the interval of $e_0$\ in the plot $(c)$ \label{fig:scattering-function-3}}
        \end{center}
        \end{figure}

%% DA COMPLETARE     The plot is shown in $(a)$ of fig. \ref{fig:scattering-function-4}.\\
      In the case of $f=f_1$\ the map links $e_0$\ to the value of $H_0$\ taken at $t^\ast$\,, after the driver is switched off (see fig. \ref{fig:scattering-function-3}):      
      \[
        S^1_{(x_0,v_0)}:\ \ e_0\ \,\mapsto\ \,H_0(p(t^\ast),\,x(t^\ast))
     \]
      For the periodic driver we introduce the one-to-many values mapping:
     \[
        S^2_{(x_0,v_0)}:\ \ e_0\ \,\mapsto\ \,H_0(p(t_k),\,x(t_k))
     \]
      $\,\nu\,t_k=\pi/2\,+\,k\,2\,\pi$, $k=1,\,2,\,\ldots $; the map links $e_0$\ to\ the energies \,$H_{0}$, every driver period. 
%$2\,\pi/\nu$.
      In all cases we obtain patterns that display some of the features of the maps investigated above:
      \begin{itemize}
       \item [-] There is a sequence of alternating regular and irregular intervals of $e_0$. 
      \item [-] A magnification of an irregular interval shows a pattern similar to the complete one (see $(b)$\ of fig. \ref{fig:scattering-function-3}). 
      \item [-] Except for very few cases  a regular interval corresponds to a determined number of crossings of $x=0$ (or number of steps in the fundamental region. See $(c)$\ of fig. \ref{fig:scattering-function-3}).
      \end{itemize} 

      We should point out that, differently from sect. $3$, the hierarchy of the regular intervals with respect to $N_c$, the number of crossings of $x=0$,\ does not completely hold. In fact, there are some regular intervals where $N_c$ changes value (see the jump\ from $N_c=5$\ to $N_c=7$\ of the regular interval at the left hand side of the plot $(d)$\ of fig.~\ref{fig:scattering-function-3} and the jump from $N_c=8$\ and $N_c=10$ of the regular interval $[4.954,4.957]$).
%This fact and our choice of the ingoing variables above suggest that the main properties of the scattering map are provided even though only the outgoing variables 
%     are asymptotic. Twisting the meaning of the word we still call it scattering function.

        We come back to the horseshoe construction that is traced out by the invariant manifolds of the outermost fixed points. We have stated that, for fixed $e_0$, the topology of homoclinic/heteroclinic
        intersections determines the structure of a scattering function corresponding to a line of initial conditions which cuts tendrils of the stable manifold close to the local branch.   
 We want to explain what occurs when we fix the initial conditions and vary  $e_0$.
%        We note that a change of $e_0$\ has the effects on the manifolds that they change their length and homoclinic bifurcations take place whenever a tip of a manifold hits another manifold.
By changing $e_0$\ the length of manifolds is affected and homoclinic bifurcations take place whenever the branches of two manifolds cross each other. 
%of manifolds h 
Accordingly the invariant 
        set changes its topology and some orbits may get lost.

\begin{figure}[ht!]
        \begin{center}
         \includegraphics[width=0.4\textwidth]{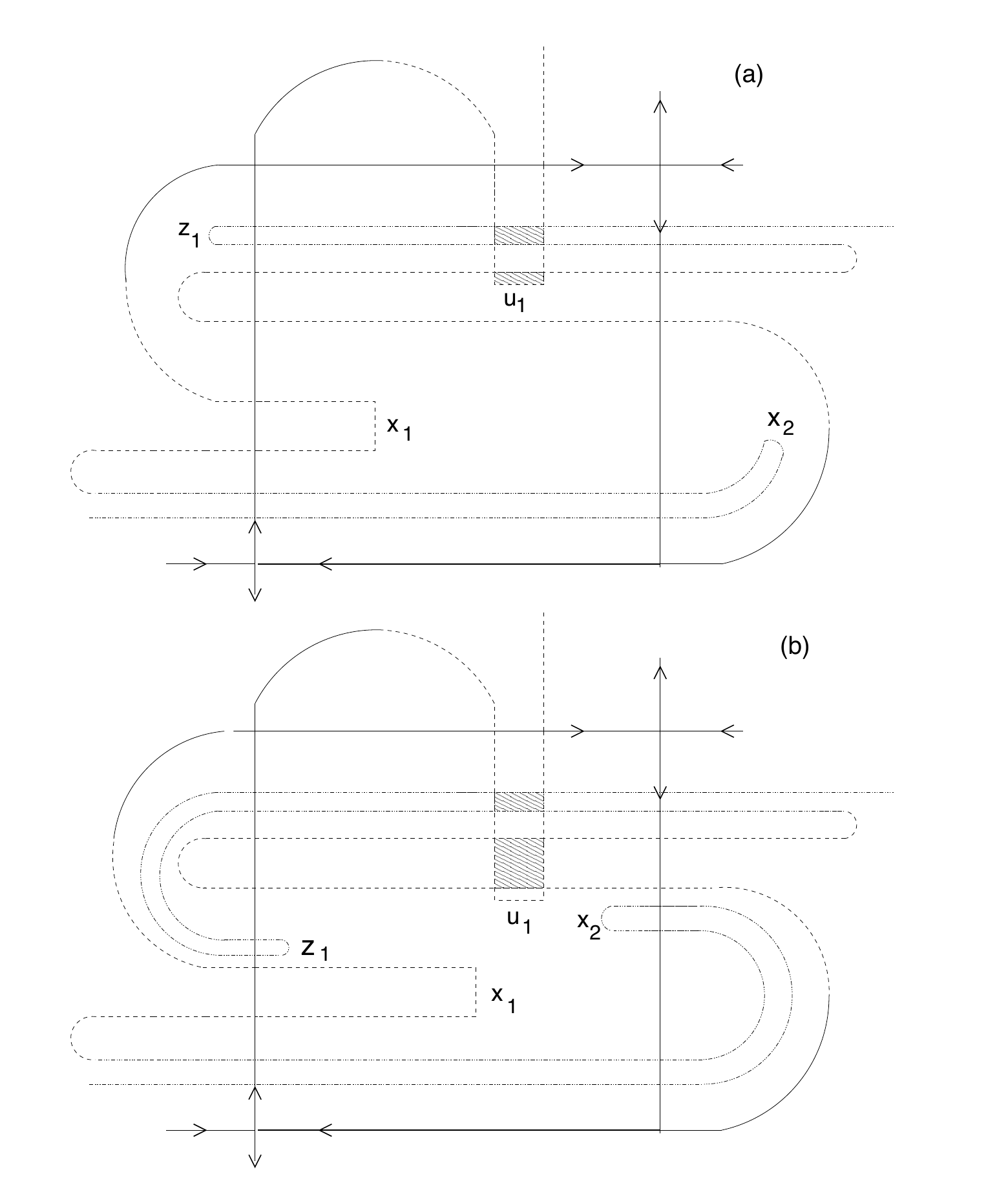}
         \caption{The plot $(a)$\ displays an incomplete horseshoe where the tips $u_1$, $x_2$\ and $z_1$ enter areas where they cannot hit invariant manifolds. In the plot (b)\ the same tips can undergo tangencies 
         under small changes of $e_0$.\label{fig:horseshoe-bifurcation}}
        \end{center}
        \end{figure}

%         If a tip riches into areas 
%        that are free of other manifolds then stretching or compression of segments of manifolds does not cause any tangency. 
        Consider $(a)$ of fig. \ref{fig:horseshoe-bifurcation}: a tip $u_1$ reaches into a stable gap that, we know, is free of further gap of $W_s$, and the  tips $z_1$\ and $x_2$\  lie in the areas of tendrils. In these situations stretching or compression of segments of manifolds do not cause any tangencies.
%There are two situations where changes of $e_0$ do not produce any bifurcation: a tip $u_1$ reaches into a stable gap that, we know, is free of further gap of $W_s$, and the  tips $z_1$\ and $x_2$\  lie in the areas of tendrils. 
        If this is the case for the tips of all gaps up to order $n$\ then, even if we vary $e_0$\ over a suitable interval, the corresponding changes of the topology may affect only orbits with at least $n$\ iterates in the fundamental region. This implies a stability of a part of the structure of orbits against bifurcations. 
%If the value of $H_0$\ refers to this part of the structure then it should depend smoothly on $e_0$.     
%%         Therefore it occurs that the topology of the horseshoe does not change if we vary $e_0$\ in an interval.\\ 
%%         We state that whenever the topology of the tangle is not modified the number of iterates in the fundamental region does not change (and $H_0$\ varies smoothly).
        On the other hand  a tip that exits a gap can undergo a homoclinic tangency for small changes of $e_0$ (see $(b)$\ of fig. \ref{fig:horseshoe-bifurcation}).
        In the fundamental region, the structure of the gaps, that is kept fixed over an interval of $e_0$, resembles the structure of homoclinic/heteroclinic intersections that is "seen" by a 
        line of initial conditions, as long as we consider orbits with a number of steps in $\mathcal{A}$ not too large. The second structure is connected to the pattern of gaps in a tendril. 
        We conclude that the structure of singularities of the maps  $ S^i_{(x_0,v_0)}$, $i=1,\,2$, should resemble to the scattering functions with respect to a line of initial conditions.

        An essential observation 
%regarding a non-hyperbolic invariant set (an incomplete horseshoe) 
        is that there are parameter intervals in which the tips of the 
        invariant manifolds do not create homoclinic bifurcations under small changes of the parameter. This occurs due to the gaps which are empty of invariant manifolds. 
        It should be pointed out that the idea of the gaps may be helpful only if it is applied to a system with an open, infinite phase space (scattering systems). In fact in a bound Hamiltonian system the invariant 
        manifolds are dense in a connected subset of the phase space not containing KAM tori.

\section{Conclusions.}
  We have studied a one-dimensional oscillator, with a single equilibrium point and critical points at infinity, driven by a force that  is either periodic or lasting a finite time interval. 
%The phase space of the system
%  is not compact.

By numerical simulations we have determined maps of  
%we build up some maps in which we measure outgoing 
scattering data (energy, delay time) and number of zeros of the orbits as functions of an initial condition (the other initial 
  condition and the driver amplitude being kept fixed) and of the driver amplitude (the initial conditions being fixed).
  The maps display a characteristic pattern. The outgoing variables are smooth functions in intervals where the number of zeros $N_c$ of the solutions does not vary; $N_c$\ marks each interval. Successive removal  of all the smooth intervals with $N_c=1,\,2,\,3,\,\dots$\ leads to a set of singularities that appears to be Cantor-like and corresponds to trajectories that never escape.
  
  The number of zeros plays the role of an order parameter that corresponds to a hierarchical structure of the regular intervals. The same ordering is provided by the delay time function. 
  This property is connected to the structure of gaps in a tendril, i.e. the area that is enclosed between a branch of the stable manifold and the local segment of the unstable manifold and that reaches out the fundamental area into 
  the incoming  asymptotic region.

  The regular intervals and the singularities pattern of the maps $S^i_{(x_0,\,\mathrm{v}_0)}$, $i=1,\,2$, are understood through the gaps in the fundamental area: some regions of phase space are void of invariant manifolds and 
  homoclinic-heteroclinic intersections cannot take place even if a change of a parameter can produce stretching or compression of manifolds.

  We introduce a map (return map) which is defined in a plane $\Gamma$\ and describes the dynamics of the crossings of $x=0$. Through the map we can identify a pattern of subsets in $\Gamma$\ whose structure is connected to the structure of the scattering map.         

  The number of orbits as a function of the time of permanence close to the origin decreases slower than expected: it decreases as a power law rather than exponentially. The explanation is due to the fact that the invariant structures are not completely destroyed: some Cantor tori remain, and a reduction in the number of orbits close to the origin is due to a stickiness effect. 
   These orbits starting far from a torus end up near it and remain there, creating correlations between different orbits and therefore decreasing the chaotic effect.

\end{document}